\documentclass[twocolumn,tighten]{aastex63}

\usepackage{amsmath}
\usepackage{comment}
\usepackage{tabularx}
\usepackage{algorithm2e}
\usepackage{float}
\usepackage{fullwidth}
\usepackage{nccmath}
\usepackage{mathtools}
\usepackage{multirow}
\usepackage{scrextend}
\usepackage{lineno}
\usepackage{rotating}

\definecolor{yscol}{rgb}{0.8, 0.6, 1}

\newcommand{\msun}{\,{\rm M}_\odot}
\newcommand{\pc}{\,{\rm pc}}
\newcommand{\kpc}{\,{\rm kpc}}
\newcommand{\myr}{\,{\rm Myr}}
\newcommand{\gyr}{\,{\rm Gyr}}

\newcommand{\s}{\,{\rm s}}
\newcommand{\g}{\,{\rm g}}
\newcommand{\cm}{\,{\rm cm}}

\newcommand{\om}{\Omega_{\rm m}}
\newcommand{\sig}{\sigma_{8}}
\newcommand{\asnone}{A_{\rm SN1}}
\newcommand{\asntwo}{A_{\rm SN2}}
\newcommand{\aagnone}{A_{\rm AGN1}}
\newcommand{\aagntwo}{A_{\rm AGN2}}

\newcommand{\bsx}{\boldsymbol{x}}
\newcommand{\bsv}{\boldsymbol{v}}
\newcommand{\bsa}{\boldsymbol{a}}
\newcommand{\Dt}{\Delta t}
\newcommand{\dt}{\mathrm{d}t}

\newcommand{\newenzo}{{\tt Enzo-N}}
\newcommand{\mcluster}{\tt Mcluster}
\newcommand{\enzo}{{\tt Enzo}}
\newcommand{\nbody}{{\tt Nbody6++GPU}}
\definecolor{yscol}{rgb}{0.8, 0.6, 1}
\def\ys#1{{\textcolor{yscol} {[#1]}}}
\def\sy#1{{\textcolor{cyan} {[#1]}}}
\def\revision#1{{{#1}}}
\def\bs#1{{{#1}}}

\newlength{\Oldarrayrulewidth}

\makeatletter
\newcommand{\thickhline}{%
    \noalign {\ifnum 0=`}\fi \hrule height 1pt
    \futurelet \reserved@a \@xhline
}
\newcommand{\thichline}{%
    \noalign {\ifnum 0=`}\fi \hrule height 0.8pt
    \futurelet \reserved@a \@xhline
}
\newcolumntype{"}{@{\hskip\tabcolsep\vrule width 0.8pt\hskip\tabcolsep}} 
\makeatother

\received{December 25, 2023}
\revised{July 15, 2024}
\accepted{August 2, 2024}
\submitjournal{ApJ}

\graphicspath{{./}{figures/}}

\begin{document}





\title{Evolution of Star Cluster Within Galaxy using Self-consistent Hybrid Hydro/N-body Simulation}

\correspondingauthor{Yongseok Jo}
\email{yj2812@columbia.edu}
\author[0000-0003-3977-1761]{Yongseok Jo}
\affiliation{Columbia Astrophysics Laboratory, Columbia University, New York, NY 10027, USA}
\affiliation{Center for Computational Astrophysics, Flatiron Institute, 162 5th Avenue, New York, NY, 10010, USA}

\author{Seoyoung Kim}
\affiliation{Center for Theoretical Physics, Department of Physics and Astronomy, Seoul National University, Seoul 08826, Korea}

\author{Ji-hoon Kim}
\affiliation{Center for Theoretical Physics, Department of Physics and Astronomy, Seoul National University, Seoul 08826, Korea}
\affiliation{Seoul National University Astronomy Research Center, Seoul 08826, Korea}

\author{Greg L. Bryan}
\affiliation{Department of Astronomy, Columbia University, New York, NY 10027, USA}





\begin{abstract}
We introduce a GPU-accelerated hybrid hydro/N-body code (\newenzo) designed to address the challenges of concurrently simulating star clusters and their parent galaxies. 
This task has been exceedingly challenging, primarily due to the considerable computational time required, which stems from the substantial scale difference between galaxies ($\sim 0.1 \mathrm{Mpc}$) and star clusters ($\sim\pc$).
Yet, this significant scale separation means that particles within star clusters perceive those outside the star cluster in a semi-stationary state.
By leveraging this aspect, we integrate the direct N-body code ({\nbody}) into the cosmological (magneto-)hydrodynamic code ({\enzo}) through the utilization of the semi-stationary background acceleration approximation.
We solve the dynamics of particles within star clusters using the direct N-body solver with regularization for few-body interactions, while evolving particles outside—--dark matter, gas, and stars--—using the particle-mesh gravity solver and hydrodynamic methods.
We demonstrate that {\newenzo} successfully simulates the co-evolution of star clusters and their parent galaxies, capturing phenomena such as core collapse of the star cluster and tidal stripping due to galactic tides.
This comprehensive framework opens up new possibilities for studying the evolution of star clusters within galaxies, offering insights that were previously inaccessible. 

\end{abstract}

\keywords{methods: numerical, galaxy: formation, galaxy: evolution, galaxies: star clusters}

\section{Introduction} 

The realm of cosmological hydrodynamic simulations has marked significant progress in enhancing our comprehension of the large-scale structure and galaxy formation. On cosmological scales, dark matter-only simulations have effectively captured the evolution of the large-scale structure \citep{springel2005MNRAS.364.1105S,klypin2011ApJ...740..102K}. Furthermore, full-physics simulations, incorporating various models for astrophysical phenomena ranging from star formation to AGN feedback, have yielded realistic cosmic galaxy populations \citep{genel2014MNRAS.445..175G,nelson2018MNRAS.475..624N,springel2018MNRAS.475..676S,ni2022MNRAS.513..670N}. Zoom-in simulations have also contributed by focusing on the evolution of galaxies within the cosmological context \citep{hopkins2010MNRAS.407.1529H,kim2019ApJ...887..120K}. Moreover, they have delved into specific areas such as the formation of the first stars \citep{abel2002Sci...295...93A,yoshida2006ApJ...652....6Y} and the formation of massive black holes in the early universe \citep{regan2017NatAs...1E..75R}.
In spite of the adaptability and versatility of cosmological simulations, offering variable resolutions and targeting diverse astrophysical phenomena, none have successfully resolved the evolution of star clusters.

A star cluster is a complex dynamical system comprising approximately a million star particles, involving intricate dynamical and astrophysical phenomena, including binaries and stellar evolution \citep{2010ARA&A..48..431P}. 
The holy grail of star cluster simulation has been to perform a self-consistent collisional N-body calculation for more than a million star particles over several billion years.
The advent of direct N-body simulations has succeeded in capturing the dynamical evolution of star clusters \citep{heggie1975MNRAS.173..729H,aasrseth1999PASP..111.1333A, makino1997ApJ...480..432M,wang2015MNRAS.450.4070W, wang2016MNRAS.458.1450W}. 
Moreover, star cluster simulations have progressed to comprehensively address not only the dynamic evolution of star clusters \citep{mcmillan1990ApJ...362..522M,heggie2006MNRAS.368..677H,hurley2012MNRAS.425.2872H} but also a diverse range of astrophysical phenomena.
This includes, but is not limited to, \revision{stellar evolution} \citep{zwart2001MNRAS.321..199P, hurley2001MNRAS.323..630H,hut2003NewA....8..337H,sippel2012MNRAS.427..167S,kamlah2022MNRAS.511.4060K,kamlah2022MNRAS.516.3266K,banerjee2020A&A...639A..41B}, X-ray binaries \citep{mapelli2013MNRAS.429.2298M, spera2015MNRAS.451.4086S,ryu2016MNRAS.456..223R}, and the formation and evolution of intermediate-mass black holes \citep{zwart2002ApJ...576..899P, zwart2004Natur.428..724P, baumgardt2004ApJ...613.1133B,arca2023MNRAS.526..429A}.

Although there have been great efforts on both fronts, achieving a self-consistent full physics simulation for star clusters on a galactic scale or larger has been challenging. 
Nonetheless, there have been endeavors to integrate hydrodynamic engines into direct N-body simulations, and vice versa.
\citet{fujii2007PASJ...59.1095F} introduces a hybrid direct-tree N-body code called {\sc BRIDGE}, aiming to simultaneously resolve both star clusters and their parent galaxies. 
It employs two different time-stepping schemes for the internal motion of star clusters and all other motions, respectively. 
The internal motion of star clusters, which generally requires higher precision, is solved by the direct Hermite integrator, while other motions are calculated with a second-order Leapfrog integrator.

Despite its sophisticated implementation, {\sc BRIDGE} lacks hydrodynamics. 
On the other hand, {\sc MUSE} and its successor {\sc AMUSE} seamlessly link between different physical modules, including various hydrodynamic engines and gravity solvers \citep{zwart2009NewA...14..369P,pelupessy2013A&A...557A..84P}.
Specifically, {\sc AMUSE} encompasses a wide range of physical modules in different domains such as ph4 (Hermite N-body), Twobody (Kepler solver), MI6 (Hermite with Post-Newtonian), MESA (Stellar evolution), Gadget-2 (TreeSPH), and more.
Another hybrid hydro/N-body code presented in \citet{hubber2013MNRAS.430.1599H} combines the smoothed particle hydrodynamics (SPH) and the N-body method in order to model the collisional star cluster in a live gas background.
The SPH gas particles are integrated with the second order Leapfrog integrator, whereas the stars are integrated with the fourth order Hermite time integrator.
More recently, the ASURA project is equipped with the SPH solver, ASURA, within the BRIDGE scheme \citep{fujii2021PASJ...73.1057F}.
For the direct integration, it employs a tree direct N-body code, {\sc PeTar} \citep{wang2020MNRAS.497..536W}.

The advent of the hybrid codes enables the study of star clusters in various astrophysical environments in a more self-consistent way.
For instance, \citet{rieder2013MNRAS.436.3695R} performs a dark-matter-only simulation for the cosmological box of $(21 h^{-1}\mathrm{Mpc})^3$, and by replacing 30 dark matter particles with star clusters of 32000 particles, investigates the cosmological evolution of star clusters in the presence of cosmological tidal fields using the AMUSE framework without hydrodynamics.
The formation and evolution of massive objects, considered as potential seeds for supermassive black holes, have been investigated through processes involving gas accretion and particle-particle collisions within a dense cluster with a mass of approximately $\sim10^4\msun$ over $\sim 10^5$ Myr \citep{boekholt2018MNRAS.476..366B,reinoso2023MNRAS.521.3553R}.
\citet{wall2019ApJ...887...62W} has performed hybrid simulations, comprising magneto-hydrodynamics, collisional gravity, and star formation and evolution, to study the formation of star clusters in turbulent gas clouds of $10^3 - 10^5 \msun$ up to $20$ Myr.


Nevertheless, these studies face limitations in terms of system size, the breadth of physics considered, and the duration of evolution.
This is primarily attributed to the strong coupling between physical modules, particularly between hydrodynamics and collisional gravity solvers.
In order to tackle this issue, we adopt a weak coupling between the `hydro' and `nbody' parts.
Here, the `nbody' part encompasses the direct N-body solver and the regularization, while the `hydro' part includes collisionless gravity, (magneto)hydrodynamics, cosmology, and sub-resolution models.
Our {\newenzo} is a novel hybrid framework designed to facilitate robust self-consistent simulations of star clusters within their parent galaxies.
This is achieved by seamlessly integrating the direct N-body code ({\nbody}) into the (magneto-)hydrodynamic code ({\enzo}).
{\nbody} is purpose-built for simulating star clusters, utilizing a direct N-body approach, fourth-order Hermite integrator, and regularizations for few-body interactions \citep{wang2015MNRAS.450.4070W}. 
On the other hand, {\enzo} is a versatile tool addressing a broad spectrum of astrophysical problems, incorporating various physics engines such as (magneto-)hydrodynamics, PM gravity solver, gas chemistry, radiative cooling, cosmological expansion, and sub-resolution models for star formation and feedback \citep{enzo2014ApJS..211...19B}.

The hydro' and nbody' parts are connected through a weak coupling facilitated by the background acceleration, which acts as a messenger enabling communication between particles undergoing direct N-body calculations and those that do not. 
The background acceleration is computed from dark matter, gas, and stars--—entities not subject to direct N-body interactions—--on the hydro' part and transmitted to the `nbody' part. 
Subsequently, the nbody' part utilizes this semi-stationary background acceleration as the tidal field, which is updated every hydro time steps. 
The exploitation of background acceleration is justified by the considerable difference in the dynamical time scales of star clusters and galaxies, approximately $1 \myr$ and $100 \myr$, respectively.
This discrepancy results in the tidal fields of the galaxy appearing semi-stationary to the particles within star clusters.
Leveraging this, our goal in {\newenzo} is to simulate the evolution of star clusters atop a semi-stationary background potential from their parent galaxy or cosmological tidal fields for more than $100 \myr$ (up to the Hubble time). 
Throughout this process, we aim to preserve the physical integrity of the star clusters using the direct N-body solver.


This paper is structured as follows:
In Sec. \ref{sec:method}, we provide a comprehensive description of the detailed implementation of {\newenzo}.
In Sec. \ref{sec:isolated_star_cluster}, we examine the efficacy of the direct N-body calculations in {\newenzo} by comparing it to {\nbody}. Additionally, we highlight the improved performance of {\newenzo} compared to {\enzo};
In Sec. \ref{sec:orbit}, we assess the robustness of background acceleration, comparing it to the analytic solution and {\enzo}.
In Sec. \ref{sec:agora_star_cluster}, we explore the scientific potential of {\newenzo} by simulating star clusters within their parent galaxies; 
Finally, we summarize our results in Sec. \ref{sec:summary}.

\begin{figure*}
    \centering
    \includegraphics[width=1.0\textwidth]{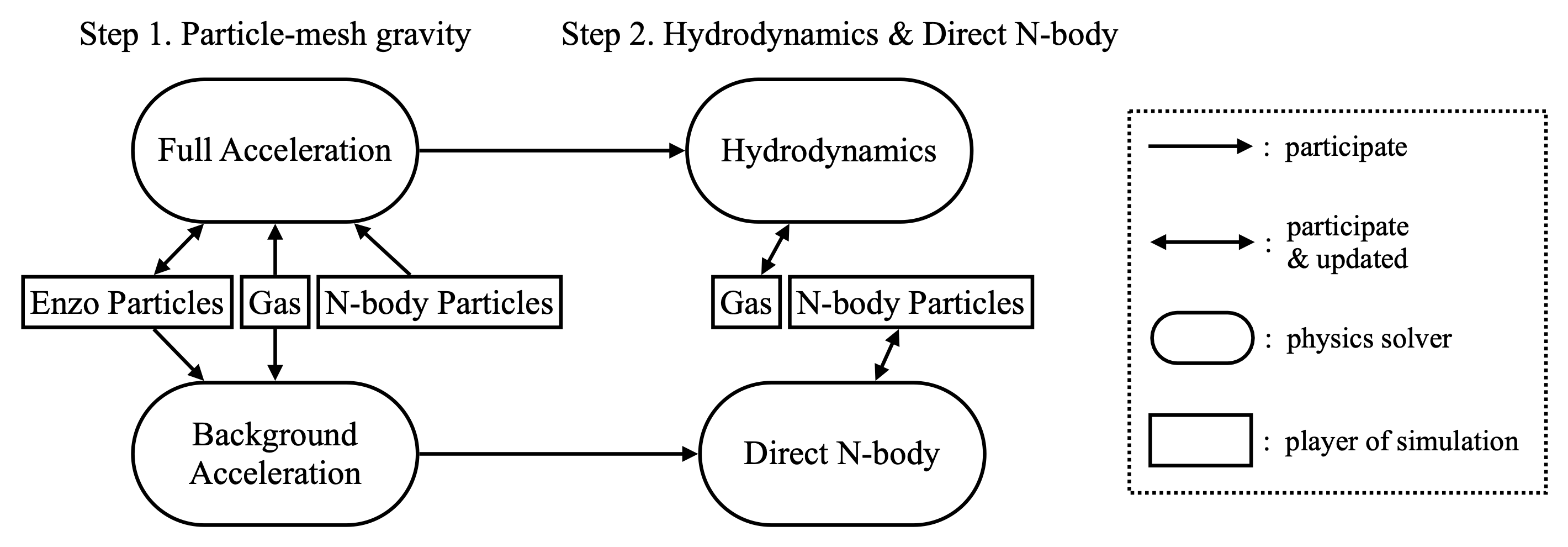}
    \caption{Overview of {\newenzo}.
   ``Enzo Particles'' encompasses entities such as dark matter, black holes, and stars, which do not play a direct role in the dynamics of star clusters.
   On the other hand, ``N-body Particles'' refers to particles, including stars and black holes, that are directly linked to the dynamics and evolution of star clusters. 
    }
    \label{fig:diagram}
\end{figure*}

\section{Numerical Method}
\label{sec:method}
In this section, we introduce {\newenzo}, a cutting-edge hybrid code that integrates a direct N-body simulation into a cosmological (magneto-)hydrodynamic code.
Through this integration, we expand the capabilities of {\nbody}, designed for simulating collisional interactions within a single or a few star clusters, to encompass galactic or cosmological scales.
Our pipeline comprises two main components: the `hydro' part and the `nbody' part. 
The hydro part, based on Enzo, handles cosmology, (magneto-)hydrodynamics, collisionless gravity, and sub-resolution physics such as radiative cooling, star formation, and black hole accretion. 
On the other hand, the nbody part is dedicated to modeling collisional gravity through the direct sum method. 
The primary role of EnzoN is to serve as the bridge connecting the hydro part and the nbody part. 
The interaction between these two components is facilitated by the use of the ``background acceleration'', which is calculated within the hydro part using the PM solver and incorporates contributions from dark matter, gas, and stars within the galaxy, excluding those in star clusters.

Fig. \ref{fig:diagram} depicts the overview of {\newenzo}.
The physical entities of {\newenzo} are categorized into two different parts: 
1. softened particles subject to particle mesh (PM) gravity solver ({\it Enzo Particles}), 
2. point mass particles subject to direct sum gravity solver that accounts for the particle-particle interaction ({\it N-body Particles}) and gas on the grids subject to hydrodynamic and PM solvers ({\it Gas}).
In the initial phase of the hydro part, we compute both the full acceleration and the background acceleration. 
In the subsequent step, the background acceleration is transmitted to the nbody part, where the dynamics of the {\it N-body Particles} is calculated using the direct sum method and regularization in addition to the background acceleration (refer to Sec. \ref{sec:method_nbody}). 
Concurrently, the hydrodynamics and subgrid physics are solved within the hydro part in parallel.

\subsection{Gravito-hydrodynamic code: {\enzo}}
\label{sec:method_hydro}
{\enzo} stands as a versatile gravito-hydrodynamic computational tool crafted to address a wide spectrum of astrophysical inquiries \citep{enzo2014ApJS..211...19B}. 
Its use extends to the realms of large-scale structure formation, galaxy evolution, and the evolution of black holes. 
It boasts a rich suite of baryon physics modules encompassing stellar processes, radiative cooling, and black hole accretion. 
{\enzo}'s hydrodynamic engine relies on an Eulerian scheme on top of block-structured adaptive mesh refinement, including four different numerical hydrodynamic methods: the hydrodynamic-only piecewise parabolic method \citep{colella1984JCoPh..54..174C,bryan1995CoPhC..89..149B}, the MUSCL-like Godunov scheme \citep{vanleer1977JCoPh..23..276V,wang2008ApJS..176..467W,wang2009ApJ...696...96W}, a constrained  transport  (CT)  staggered  MHD  scheme  \citep{collins2010ApJS..186..308C}, and the second-order finite difference hydrodynamics method \citep{stone1992ApJS...80..753S,stone1992ApJS...80..791S}.
Concurrently, gravity is addressed through the implementation of the Particle Mesh solver (PM solver), along with the mesh structure. 
The collisionless particles governed by the PM solver are advanced through a single timestep using a drift-kick-drift algorithm \citep{hockney1988csup.book.....H}.
This ensures second-order accuracy even in the presence of varying time-steps due to dynamical grid structures of adaptive refinement.

\subsubsection{Time-stepping in {\enzo}}
{\enzo} adaptively integrates  the  equations, owing to the adaptive mesh-grid structure, not only in  space  but also in time.
The time step is determined on a level-by-level basis, ensuring the selection of the largest timestep that satisfies various criteria for different physical components. 
These criteria encompass hydrodynamics, acceleration, particle velocity, radiation pressure, heat conduction, and the expansion of the universe (refer to Sec. 9 in \citet{enzo2014ApJS..211...19B} for details).
For instance, the time step for hydrodynamics is given as 
\begin{equation}
    \Delta t_{\rm hydro} = \min\left(\kappa_{\rm hydro}\left(\sum_{x,y,z}\frac{c_{\rm s}+|v_{x}|}{a\Delta x}\right)^{-1}\right)_{\rm L},
\end{equation}
where $c_s$, $a$, $\Delta x$, and $\kappa$ are, respectively, the sound speed, cosmic scale factor, cell size, and a dimensionless constant for ensuring that the Courant–Freidrichs–Levy condition is always met.
Here, $\min(\cdots)_\mathrm{L}$ finds the minimum value for all cells or particles on a given level L.
In this work, we adopt $\kappa_{\rm hydro}$ of 0.3 (typically set to $0.3-0.5$).
In the case of acceleration, the time step is determined by
\begin{equation}
    \Delta t_{\rm acc} = \min\left(\sqrt{\frac{\Delta x}{|\boldsymbol{g}|}}\right)_{\rm L}.
\end{equation}
The combination of the time step and drift-kick-drift algorithm ensures second-order accuracy for gravity \citep{hockney1988csup.book.....H}.

\subsection{Direct N-body code: {\nbody}}
\label{sec:method_nbody}

We present here descriptions of algorithms used in {\nbody} \citep{wang2015MNRAS.450.4070W}. Most methods implemented in {\nbody} are retained in {\newenzo}, except for methods that needs to commensurate with {\enzo}, such as time steps.

\subsubsection{Hermite integration with individual time steps}
\label{sec:method_hermite}

{\nbody} is a direct {\it N}-body code based on a fourth-order Hermite integration method \citep{makino1991ApJ...369..200M}. This method utilizes the lower order derivatives to obtain approximate values for higher order derivatives of each particle. Block time steps are assigned to each particle so that the quantized time step for each particle is smaller than the time step criteria. The criteria of time steps for the particles are given by

\begin{equation}
\label{eq:tcrit}
    \Delta t = \sqrt{\eta \frac{ |\mathbf{a}^{(0)}| |\mathbf{a}^{(2)}| + |\mathbf{a}^{(1)}|^{2} }{ |\mathbf{a}^{(1)}| |\mathbf{a}^{(3)}| + |\mathbf{a}^{(2)}|^{2} } },
\end{equation}
where $\Delta t$ is the criteria for time step and $\mathbf{a}^{(n)}$ is the n-th derivative of acceleration \citep{aarseth2006GReGr..38..983M}. The steps are quantized in powers of 1/2, where the unit for time step is Hénon units \citep{heggie1986LNP...267..233H}. 
The time-stepping is independently managed within the 'hydro' and 'nbody' parts. 
Communication is initiated when $n\Delta t_\mathrm{nbody} > \Delta t_\mathrm{hydro}$ for an arbitrary number of time steps $n$, where $\Delta t_\mathrm{nbody}$ is the block time step of the `nbody' part, and $\Delta t_\mathrm{hydro}$ is the finest time step of the 'hydro' part. 
During communication, the velocities and positions of N-body particles are transmitted to the 'hydro' part, and the background acceleration is sent to the 'nbody' part. 
However, in case of a mismatch between the time steps of the 'nbody' and 'hydro' parts, the positions and velocities of particles within the 'nbody' part are extrapolated to match the time step of the 'hydro' part.
The calculation in the `nbody' part resumes from the extrapolated time in blocks with quantized values of Hénon units. 
The time steps used for extrapolation are set to be equal or smaller than the time step given by Eq. \ref{eq:tcrit}. 
Furthermore, the time steps of the `nbody' part retain their consistency regardless of communication with the `hydro' part, ensuring that the physics remains unaffected. 
The detailed correspondence between {\nbody} and {\newenzo} in terms of physical results is discussed in Sec. \ref{sec:isolated_star_cluster}.

\subsubsection{Ahmad-Cohen neighbor scheme}
\label{sec:method_ac-scheme}

Direct calculation of particles up to fourth-order is computationally expensive, so {\nbody} employs a neighbor scheme described by \citep{ahmad1973JCoPh..12..389A}. In this scheme, the total force on a particle is divided into short-range ({\it irregular}) and long-range ({\it regular}) components. The irregular force is calculated from neighbor particles within a certain radius of the target particle. All the other particles are considered as contributors for regular force. The irregular force is updated at shorter intervals than the regular force. During irregular time steps, the regular force is estimated up to first-order using the derivatives calculated at regular time steps. In {\newenzo}, an additional force component from {\enzo} must be considered as well. Since the fluctuation of the background force from {\enzo} is expected to be less intensive than {\nbody}, the background force component is added to the regular force components when calculating.

\subsubsection{Regularization of close encounters}
\label{sec:method_regularization}

Accurate modeling of close encounters is crucial for the correct evolution of star clusters; 
however, handling close encounters using the Hermite scheme is both time-consuming and inaccurate.
Therefore, algorithms of \citep{KustaanheimoSCHINZELDAVENPORTSTIEFEL+1965+204+219} and \citep{mikkola1993CeMDA..57..439M} are adopted in {\nbody}.
Particles in binary or triple systems are taken out and replaced by their center of mass in the main integration code. 
The relative positions of members are calculated using regularization algorithms, while the center of mass is updated by the Hermite scheme.

\subsection{Center of Mass Frame}
\label{sec:method_com}
Simulating star clusters on a larger scale poses a challenge due to the hindrance caused by the peculiar or bulk motion of the star clusters, which significantly hampers computational speed. 
As discussed in Sec. \ref{sec:method_hermite}, the time-stepping in the `nbody' part is determined as a function of acceleration and its derivatives.
While the velocity dispersion of a star cluster is on the order of 10 km/s, that of a Milky-Way mased galaxy is on the order of $10^2$ km/s, indicating the relative magnitudes of self-gravity.
In the case of a star cluster situated in the spiral arms, the gravitational force exerted on the center of mass of the star cluster significantly surpasses its self-gravity.
\revision{
This may lead to issues, primarily regarding time steps.
In Eq. \ref{eq:tcrit}, we use acceleration and its derivatives to determine time steps for each particle.
However, our current setup cannot obtain higher-order derivatives for tidal accelerations from the background, potentially resulting in unphysical time steps.
This problem exacerbates when tidal accelerations that affect the motion of the center of mass dominate the accelerations of particles processed in the 'nbody' part.
}
To mitigate this issue, we routinely update the positions, velocities, and background acceleration of the N-body particles in the center of mass frame within the 'nbody' part, while analytically solving for the bulk motion.
\revision{This results in only non-bulk tidal gravity remaining, which is generally much weaker than the self-gravity of star clusters.}

\begin{figure}
    \centering
    \includegraphics[width=0.47\textwidth]{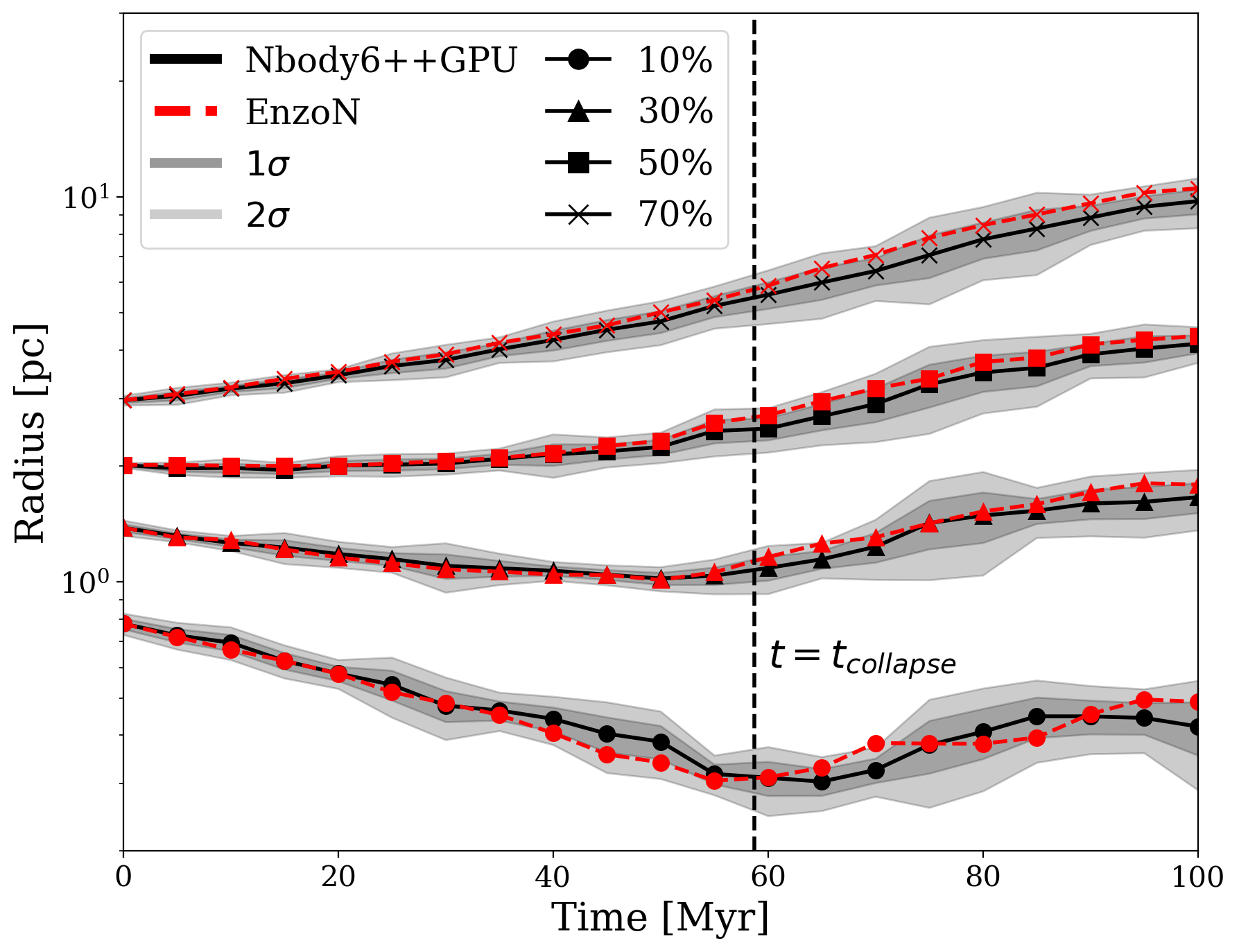}
    \caption{Lagrangian Radius evolution of clusters with Plummer profile simulated in {\nbody} ({\it black solid}) and {\newenzo} ({\it red dashed}). The {\it shaded grey} regions indicate the 1 $\sigma$ and 2 $\sigma$ values of 8 {\nbody} simulations. From {\it bottom} to {\it top}, radius enclosing 10\%, 30\%, 50\%, 70\% mass of the cluster is shown. The horizontal line ({\it black dashed}) indicate the estimated core collapse time for the simulated cluster. The radius evolution of {\nbody} and {\newenzo} agrees well with each other. 
    }
    \label{fig:lagr_nb_enzon}
\end{figure}

\section{Isolated Star Cluster}
\label{sec:isolated_star_cluster}
In this section, we conduct a series of tests using various idealized clusters to evaluate the robustness and efficiency of the collisional gravity solver integrated into {\newenzo} by comparing it against the performance of {\nbody} and {\enzo}. All the initial conditions needed for {\nbody} are generated based on the code {\mcluster} \citep{kupper2011MNRAS.417.2300K}. Options for {\nbody} using a tidal field, stellar evolution, and removal of escapers are suppressed so that fair comparison between the two simulations can be made.
The star cluster generated has a total mass of $10^5 \msun$ and a half-mass radius of 2 pc and is composed of 1000 particles with equal mass. The positions and velocities of the stars follow a homogeneous Plummer profile defined as
\begin{equation}
\rho(r)=\frac{3M_0}{4\pi a^3}\left(1+\frac{r^2}{a^2}\right)^{-5/2}
\end{equation}
where $a=1.533$ \revision{$\bs{\pc}$} and $M_0=10^5\msun$. Two types of models were generated using the conditions above, one without primordial binaries and one with 500 primordial binaries. The primordial binaries follow period and eccentricity distributions given from \citep{kroupa1995MNRAS.277.1507K}. 
We employ the star cluster without any primordial binaries throughout the paper unless otherwise noted. 
Lastly, we set the spatial resolution of {\newenzo} to be notably low ($\sim\kpc$) throughout this section as we do not use the 'hydro' part.

\begin{figure}
    \centering
    \includegraphics[width=0.47\textwidth]{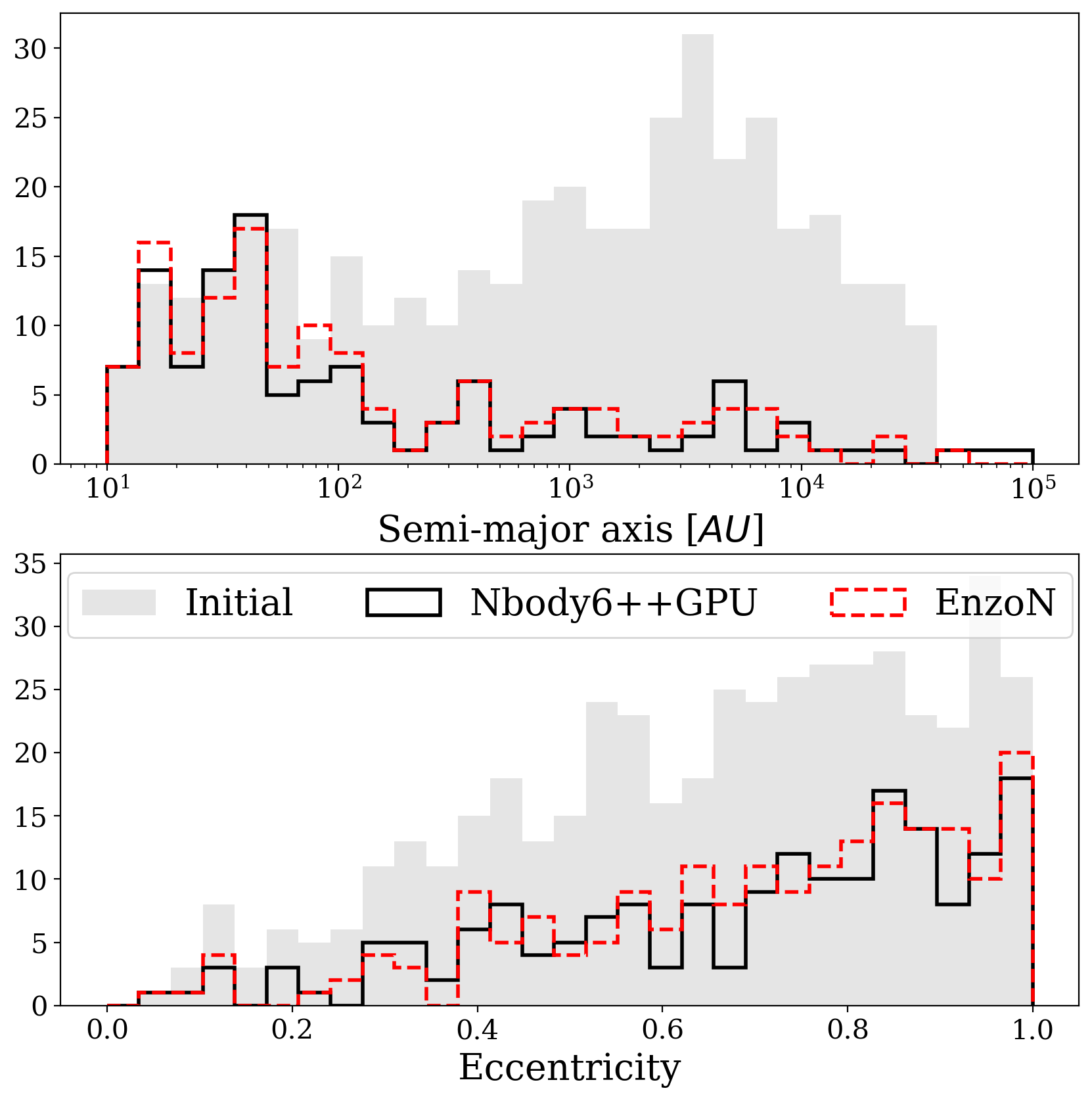}
    \caption{Histogram of semi-major axis ({\it top}) and eccentricity ({\it bottom}) of binaries in the star cluster. The {\it grey shaded} region represents the initial distribution of binaries, whereas the lines represents distribution of binary properties in {\nbody} ({\it black solid}) and {\newenzo} ({\it red dashed}) after 100 Myr. Overall, the two distributions closely align with each other.}
    \label{fig:bin_hist}
\end{figure}

\subsection{{\newenzo} versus Nbody6++}
\label{sec:iso_enzon_nbody6}

To test the robustness of {\newenzo} at simulating star clusters, we first compare the results of a long term evolution of the Plummer model of {\newenzo} with {\nbody}. The total simulation time is set to 100 Myr, enough to exceed the expected core collapse times for the clusters used. We perform eight independent shadow simulations for both {\newenzo} and {\nbody} using initial conditions generated from different random seeds, while having the identical star cluster profile. The results shown are means and standard deviations obtained from the eight shadow simulations. The distance between the star particle and its 6th nearest neighbor is used as the density estimator for calculating the density center (refer to equation II.3 and V.3 provided in \citet{casertano1985ApJ...298...80C}). The Lagrangian radii are then calculated based on the distance of stars from the cluster's density center. 

The evolution of Lagrangian radii corresponding to 10\%, 30\%, 50\%, 70\% of the total mass (from {\it bottom} to {\it top}) in Plummer model for {\nbody} ({\it black solid}) and {\newenzo} ({\it red dashed}) is presented in Fig. \ref{fig:lagr_nb_enzon}. The two results exhibit strong agreement, indicating that {\newenzo} can successfully model the evolution of star clusters. Moreover, the core collapse occurs at $\sim 60 \myr$, which is in accordance with the estimated core collapse time of $\sim 15\, t_\mathrm{rh}$ \citep{spitzer1975ApJ...200..339S}. $t_\mathrm{rh}$ is the half-mass relaxation time given by $t_\mathrm{rh}= (0.138/\ln{\Lambda})(Nr_{h}^3/Gm)^{1/2}$, where $r_{h}$ is the half-mass radius and $\Lambda$ is the Coulomb logarithm \citep{spitzer1988degc.book.....S}. 

In order to validate whether {\newenzo} is able to resolve sub-resolution physics such as the evolution of binaries, we performed additional tests using the Plummer star cluster with 500 primordial binaries. Fig. \ref{fig:bin_hist} shows the initial ({\it grey shaded}) and final distribution of binary properties for {\nbody} ({\it black solid}) and {\newenzo} ({\it red dashed}).
Similar to the results of Lagrangian radii, the histogram of two simulations exhibit good correspondence even after 100 Myr. We therefore confirm the ability of {\newenzo} to resolve physical problems at multiple scales, ranging from scales of star clusters (approximately pc) to binaries (approximately au).

\begin{figure}
    \centering
    \includegraphics[width=0.47\textwidth]{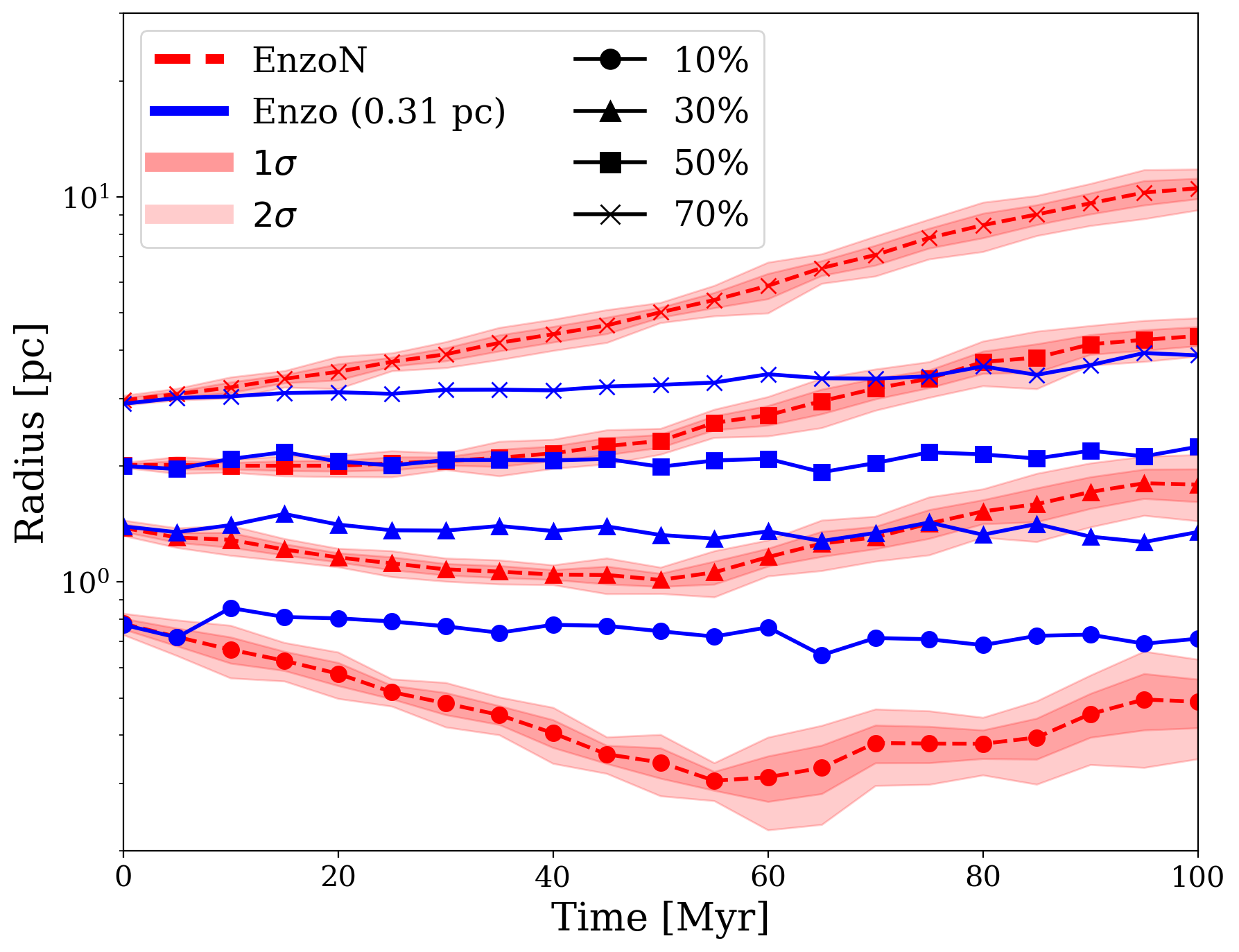}
    \caption{Lagrangian Radius evolution of clusters with Plummer profile simulated in {\newenzo} ({\it red dashed}) and {\enzo} ({\it blue dotted}). The {\it shaded red} regions indicate the 1 $\sigma$ and 2 $\sigma$ values of 8 {\newenzo} simulations. From {\it bottom} to {\it top}, radius enclosing 10\%, 30\%, 50\%, 70\% mass of the cluster is shown. There is no significant evolution of radius with respect to time in {\enzo} simulations. 
    }
    \label{fig:lagrangian_radius_enzo_enzon}
\end{figure}
\subsection{{\newenzo} versus Original {\enzo}}
\label{sec:iso_enzon_enzo}

We also compare the evolution of the isolated Plummer model clusters in {\newenzo} and {\enzo} ({\it blue dotted}). In the simulations performed with {\newenzo}, the dimensions of the root grid are $64^{3}$, and there are four static levels of refinement, leading to a cell size of 1.2 kpc for the finest resolution. The stellar particles are assigned to N-body particles, and thus calculated separately when calculating their evolution. For simulations with {\enzo}, there are nine static levels of refinement, and the particles are dynamically refined up to 16 level, leading to a resolution of 0.31 pc for stellar particles.

Fig. \ref{fig:lagrangian_radius_enzo_enzon} shows the evolution of Lagrangian radius for {\newenzo}  ({\it red dashed}) and {\enzo} ({\it blue dotted}). Clearly, the change of the Lagrangian radii in the two simulations show a significant difference. The cluster of {\enzo} retains a condition similar to its initial conditions, and does not show characteristics expected for isolated star clusters such as the core collapse. Failure of {\enzo} to reproduce the evolutionary behaviours of clusters even at high resolution can be attributed to the fact that consideration of few-body encounters are crucial for the analysis of the star cluster evolution \citep{spitzer1988degc.book.....S}. Hydrodynamics simulations like {\enzo} treats particles in a single cell as a group, and thus is an inappropriate tool for capturing such effects.
Moreover, the computational time required for {\enzo} is more than tenfold greater compared to {\newenzo}.

\begin{figure}
    \centering
    \includegraphics[width=0.47\textwidth]{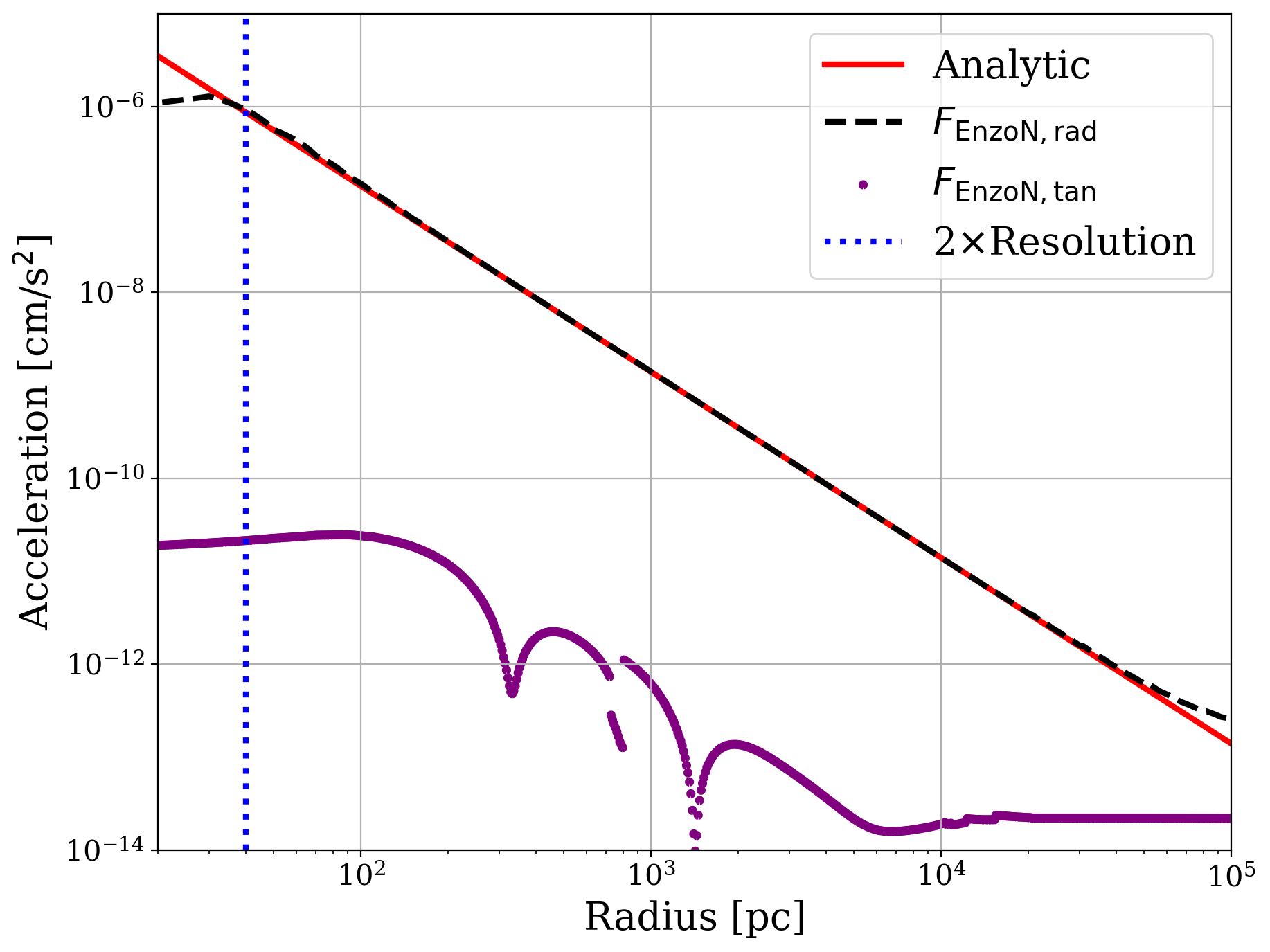}
    \caption{
    Background acceleration on the test particles as a function of the distance to the massive particle in {\newenzo}, of which radial component and tangential component are plotted in a {\it black dashed} line and {\it purple dots}, respectively. 
    The analytic solution is given in ({\it red solid}).
    The massive particle of $10^8\msun$ is situated at the center of the simulation, with test particles of $10^{-12}\msun$ evenly distributed around it.
    The resolution limit at which the PM gravity solver's effectiveness diminishes corresponds to twice the size of the finest cells ({\it blue dotted}).
      }
    \label{fig:background_analytic}
\end{figure}

\section{Orbiting Isolated Star Cluster Around Massive Point Particle}
\label{sec:orbit}
In this section, we focus on examining the stability and reliability of the background gravity implemented in {\newenzo}, which influences the N-body particles on top of the direct N-body calculations.
The background gravity is computed based on the contributions from dark matter, stars, and gas, and exerts an influence on the N-body particles within the star clusters.
This is the single most important component that links between {\enzo} and {\nbody}. 
Note that the hydrodynamic solver in {\newenzo} and {\enzo} is not included throughout this section.

\begin{figure}
    \centering
    \includegraphics[width=0.47\textwidth]{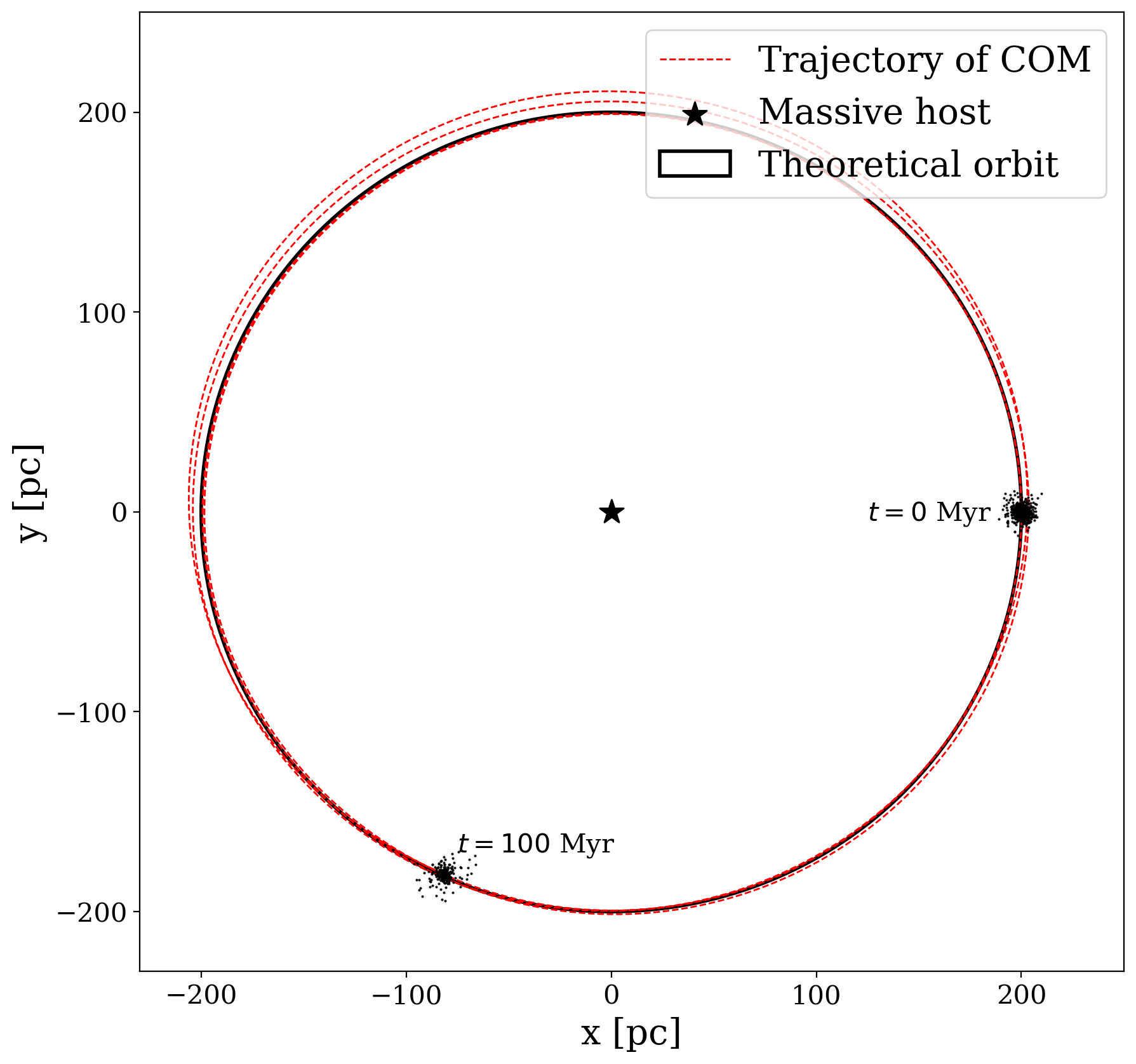}
    \caption{Trajectory of the center of mass (COM) of the star cluster of $10^5\msun$ orbiting around a massive point particle of $10^8\msun$ ({\it black star}). 
    The trajectories of {\newenzo} {\it red cross} and the analytic solution ({\it black dashed circle}) closely align with each other, exhibiting minimal deviation. 
    For a comprehensive analysis of this error, please refer to Fig. \ref{fig:orbit_r_theta}.
    }
    \label{fig:orbit_trajectory}
\end{figure}

\subsection{Comparison to Analytic Solution}
\label{sec:orbit_analytic}
We first compute the background acceleration of nearly massless particles in a gravitational field generated by a massive point particle. 
The dimensions of the root grid are $64^3$, and there are nine static levels of refinement, which leads to the finest spatial resolution of $20 \pc$.
We place a massive point particle of $10^8\msun$ at the center, while test particles of $10^{-12}\msun$ are distributed uniformly around the central particle.
The test particles are assigned the N-body particle type for gravity calculations, whereas the central particle is set to be dark matter. 
We then extract the background acceleration acting on the test particles.

Fig. \ref{fig:background_analytic} shows the radial ({\it black dashed}) and tangential ({\it purple dots}) components of the background acceleration on the test particles as a function of the distance to the central massive particle. 
The radial background acceleration computed in {\newenzo} ({\it black dashed}) is completely on top of the analytic result ({\it red solid}) $F_\mathrm{rad}\propto r^{-2}$ except for the extremes of radius.
The lower end is due to the resolution limit which is indicated by the {\it blue dotted} line, and the higher end is because of periodicity of the simulation box. 
Although the tangential component should be zero in principle, the relative error is generally less than $10^{-3}$ (refer to Fig. 10 in \citet{enzo2014ApJS..211...19B} for comparison).
Overall, we conclude that implementation of the background acceleration is robust.

\begin{figure}
    \centering
    \includegraphics[width=0.47\textwidth]{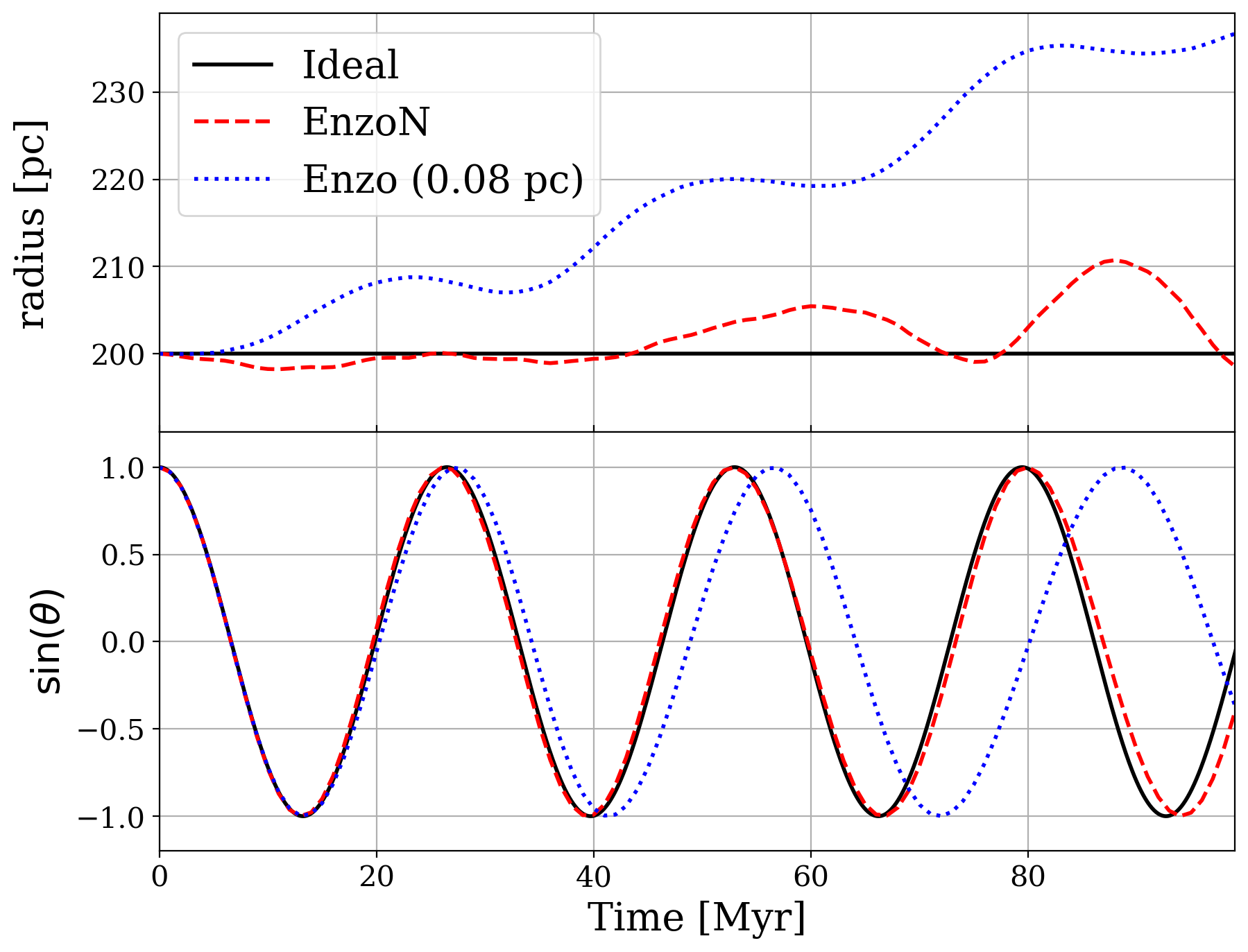}
    \caption{
    Radial and angular traces of the star cluster of {\newenzo} ({\it red dashed}) and high resolution (0.08 pc) {\enzo} ({\it blue dotted}) in comparison to the analytical solution ({\it black solid}).
    The relative errors of the radial traces averaged over 100 Myr are $1.32\%$ and $8.91\%$ for {\newenzo} and `high-res' {\enzo}, respectively.
    }
    \label{fig:orbit_r_theta}
\end{figure}

\subsection{Star Cluster Orbiting around Massive Particle}
\label{sec:orbit_circular}
We now shift our focus to the interaction between the background acceleration and the direct N-body solver.
We put a star cluster of $10^5\msun$ with 1000 equal-mass particles on a circular orbit around a massive point particle of $10^8\msun$.
The radius and period of the orbit are $200 \pc$ and $26.6 \myr$, respetively.
We have adopted a refinement scheme that uniformly refines a $(1.2 \kpc)^3$ box that encompasses the entire orbits completely, with a cell size of $20 \pc$.
We have run the simulations for $100 \myr$, which corresponds to $\sim 3.8$ revolutions.

\begin{figure}
    \centering
    \includegraphics[width=0.47\textwidth]{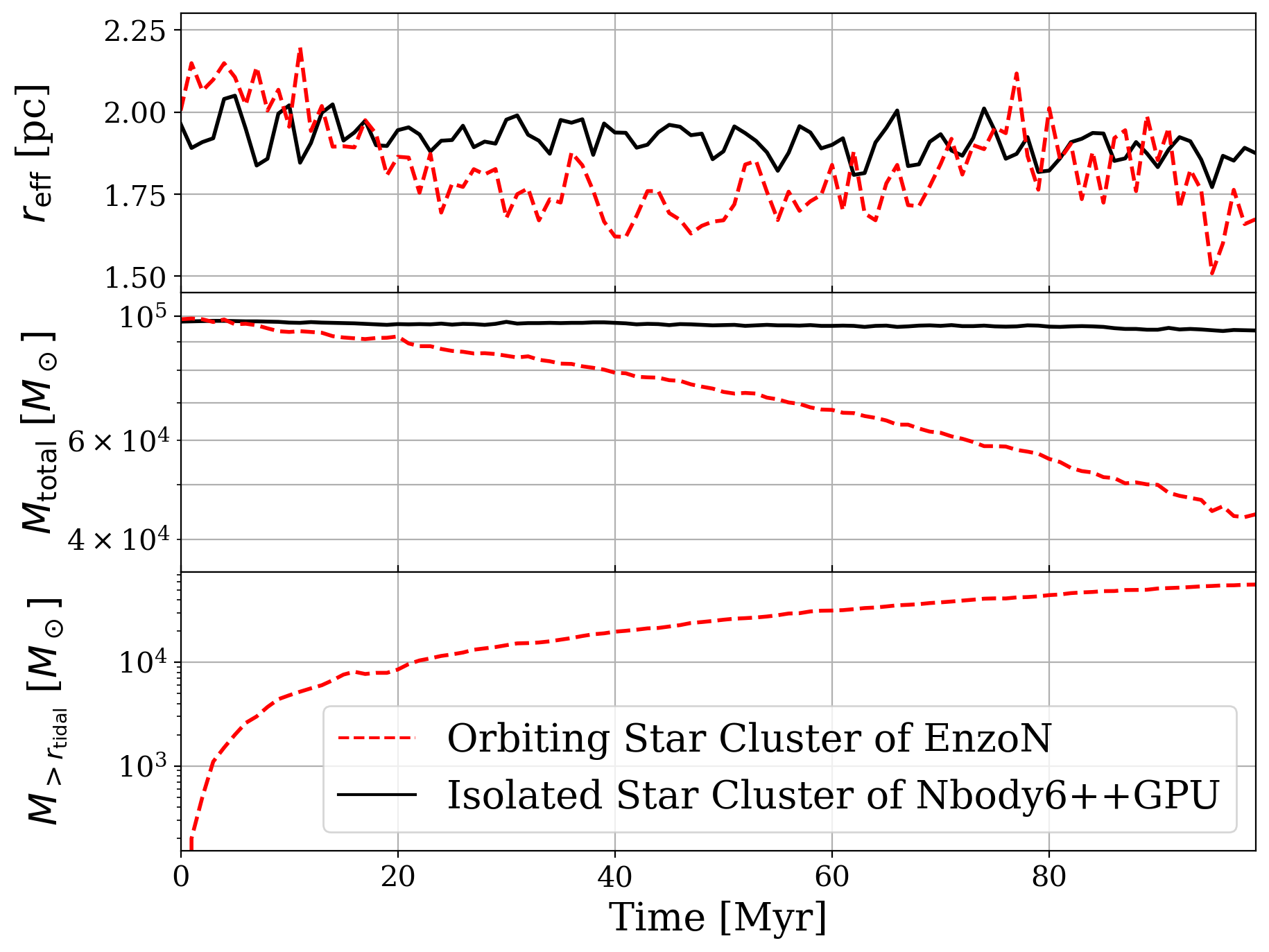}
    \caption{
    {\it Top}: Effective radius of the orbiting star cluster in {\newenzo} ({\it red dashed}) and the isolated star cluster in {\nbody} ({\it black solid}) as a function of time.
    {\it Middle}: Total mass of the clusters as a function of time.
    The total mass is defined as the enclosed mass within a radius at which the surface density equals the threshold value of $1 \msun/\pc$.
    {\it Bottom}: Mass loss owing to tidal stripping in time.
    The tidally stripped mass is calculated as the sum of the particles outside the tidal radius $r_\mathrm{tidal}=(M_\mathrm{cluster}/\revision{\bs{3}}\,M_\mathrm{host})^{1/3}R$ where $R$ is the orbital radius.
    }
    \label{fig:orbit_mass_radius}
\end{figure}

Shown in Fig. \ref{fig:orbit_trajectory} are the trajectories of the theoretical prediction ({\it black solid}) and the center of mass of the star cluster ({\it red dashed}) orbiting around the massive point particle ({\it star shape}). 
The trajectory followed by the star cluster closely aligns with the theoretical orbit, although some deviations become particularly noticeable in the upper-left section of the orbit.
A discussion of the error is presented in Fig. \ref{fig:orbit_r_theta}.
Exhibited as {\it black dots} are the star cluster particles located within a 20 pc radius at two different time points, at $0 \myr$ and $100 \myr$.
At $100 \myr$, a fraction of the particles becomes dispersed randomly throughout the simulation as a consequence of tidal stripping, as discussed in Fig. \ref{fig:orbit_mass_radius} and at the end of this section.
Nonetheless, the overall structure of the star cluster seemingly remains intact and has not suffered destruction.

We further scrutinize the reliability of this result by comparing it to the high resolution original {\enzo} run.
We set up an identical orbit simulation in {\enzo}, but with an intensive refinement scheme that uniformly refines a $(1.2 \kpc)^3$ box with a cell size of $0.08 \pc$, which is $2^8$ times finer than the {\newenzo} run. 
Fig. \ref{fig:orbit_r_theta} illustrates the radial ({\it top}) and angular ({\it bottom}) positions of the star cluster in {\newenzo} ({\it red dashed}) and {\enzo} ({\it blue dotted}) over time. 
The orbital radius of {\enzo} deviates notably from the ideal trajectory ({\it black solid}), exhibiting an average percentage error of 8.91\% over 100 Myr. 
In contrast, the orbital radius of {\newenzo} aligns more closely with the ideal path, with an average percentage error of only 1.32\%.
Examining the angular position reveals periodic variations. 
{\newenzo} deviates by less than 1 Myr from the ideal period of $\sim 26.6 \myr$, while {\enzo} exhibits a deviation of approximately $\sim 10 \myr$ over the course of $100 \myr$.
It is remarkable to note that {\enzo} is run with a spatial resolution of $0.08 \pc$, whereas {\newenzo} is performed with $20 \pc$. 
This results in a considerable disparity in computational time, with {\newenzo} demonstrating an impressive eleven fold speed advantage over {\enzo}.
The key takeaway is that {\newenzo} outperforms {\enzo} even when {\enzo} is executed with significantly higher resolution.

Up to this point, we have observed that our framework exhibits improved performance compared to the original {\enzo} while preserving the structural integrity of the star cluster. 
Lastly, we delve into the detailed evolution of the orbiting star cluster to determine if {\newenzo} can indeed achieve the best of both worlds from {\enzo} and {\nbody}.
In contrast to an isolated star cluster, the one in orbit is subject to the gravitational influence of the massive object, typically resulting in tidal phenomena such as tidal stripping caused by the gradient of gravity from the primary exerted onto the secondary. 
Fig. \ref{fig:orbit_mass_radius} illustrates the evolution of the star cluster in terms of effective radius ({\it top}), total mass ({\it middle}), and mass loss due to tidal stripping ({\it bottom}).
Here, the total mass of the cluster, denoted as $M_\mathrm{total}$, is defined as the enclosed mass within a radius where the surface density equals the threshold value of $1\msun/\pc^2$. 
The effective radius $r_\mathrm{eff}$ is determined as the half-mass radius based on $M_\mathrm{total}$.
Lastly, the mass loss is the sum of the mass of the particles located outside the tidal radius.
The tidal radius, defined as the radius at which the gravitational force from the host equals the self-gravity of the star cluster, determining the boundary beyond which particles can be torn apart, is calculated as $r_\mathrm{tidal}=(M_\mathrm{total}/\revision{\bs{(3+e)}}\,M_\mathrm{host})^{1/3}R$ where $R$, $M_{\rm host}$, and \revision{eccentricity $\bs{e}$} are $200 \pc$, $10^8\msun$, \revision{and 0 for the circular orbit} \citep{king1962AJ.....67..471K}.
In this case, the tidal radius is \revision{${\bf \sim14\bs{\pc}}$ at 0 $\bs{\myr}$}.
In comparison to the isolated star cluster ({\it black solid}) of {\nbody}, which exhibits a relatively stable evolutionary history, the star cluster orbiting around the massive particle ({\it red dashed}) undergoes a decline in the total mass and shrinks in size, coinciding  with an increase in mass loss due to tidal stripping.
The mass loss and the decrease in the total mass are distinctly comparable, despite being obtained through entirely independent methods.
This leads to the conclusion that {\newenzo} can robustly simulate a star cluster in the presence of background acceleration while preserving the structural integrity of the star cluster. 
Moreover, the secondary effects from the background acceleration, such as tidal stripping, can be resolved within a fraction of the time that {\enzo} requires, and with transcendently higher precision.

\begin{figure*}
    \centering
    \includegraphics[width=1.0\textwidth]{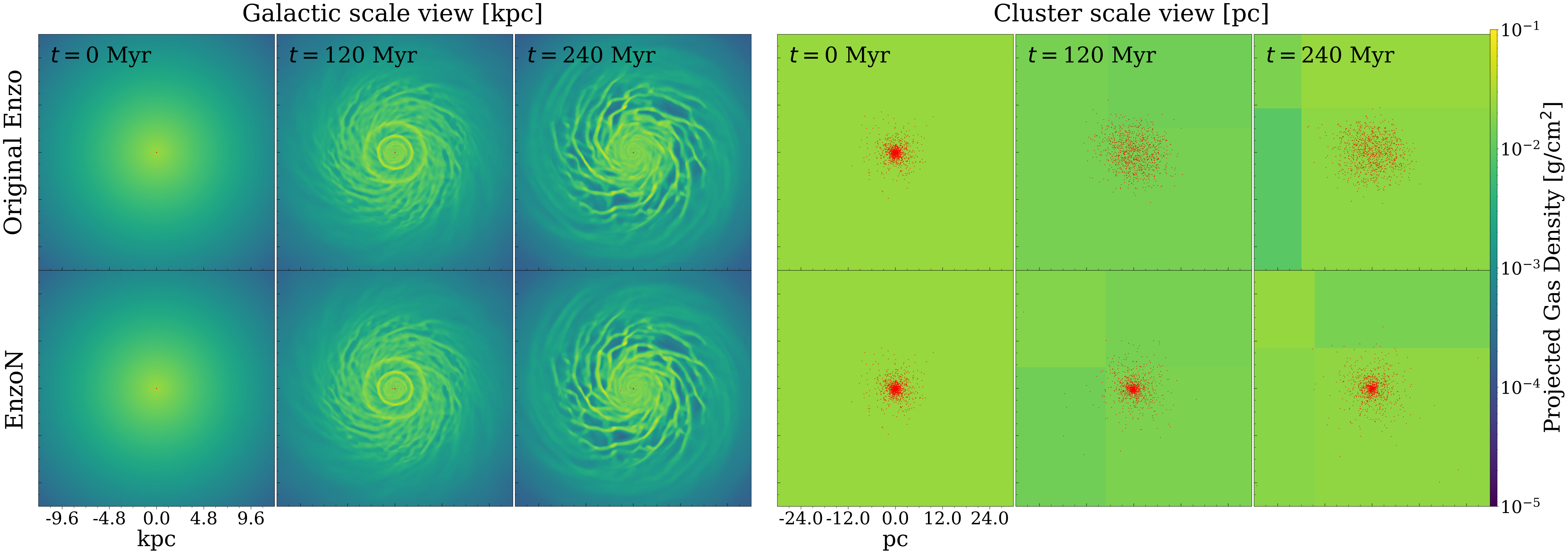}
    \caption{Time series of gas projections of the isolated galaxy with the nuclear star cluster.
    Both {\enzo} ({\it top}) and {\newenzo} ({\it bottom}) begin with the identical initial condition ({\it left most}), as provided by the AGORA project \citep{kim2016ApJ...833..202K}, which eventually evolves into a Milky Way-mass galaxy.
    Shown are the gas projections ({\it right panels}) and the zoom-in projections ({\it left panels}) with the nuclear star clusters ({\it red dots}).
    Significantly, the nuclear star cluster in {\newenzo} remains structurally intact, while the one in {\enzo} is disassembled. 
    Despite the discrepancy in the nuclear star clusters, the evolutionary patterns of the gas density in {\enzo} and {\newenzo} are indisputably identical. 
    }
    \label{fig:agora_nsc}
\end{figure*}

\section{Star cluster within a galaxy}
\label{sec:agora_star_cluster}
Thus far, we have observed that our new framework {\newenzo} outperforms the original {\enzo} while maintaining the shape of the star cluster through the integration of sophisticated gravity from {\nbody}. 
Now, we direct our attention to milestone tests for the synergetic applications. 
In this section, we test how robustly the star cluster can evolve in intricate environments such as within a galaxy.
To this end, we implant the star cluster of $10^5 \msun$ with 1000 particles into the Milky Way-mass galaxy produced by the AGORA project \citep{kim2016ApJ...833..202K}.
The initial condition of the AGORA galaxy has the following components: 
1. a dark matter halo with $M_{200}=10^{12}\msun$ and $R_{200}=205.5\kpc$ following the Navarro-Frenk-White (NFW) profile with concentration parameter of 0.01 and spin parameter of 0.04,
2. an exponential disk of $4.297\times10^{10}\msun$ with scale length of $3.432 \kpc$ and scale height of $0.3432 \kpc$, which is composed of 80\% stars and 20\% gas, and
3. a stellar bulge of $4.297\times10^9\msun$ following the Hernquist profile.
However, star formation and feedback are not implemented in this work.
We perform simulations for two distinct scenarios, varying the initial placement of the star cluster.
In Section \ref{sec:agora_nsc}, the nuclear star cluster is positioned at the center of the galaxy, whereas in Section \ref{sec:agora_outskirts}, the star cluster is situated on the outskirts of the galaxy at a distance of $x=8\kpc$ from its center.

\subsection{Nuclear Star Cluster in the Center of the Galaxy}
\label{sec:agora_nsc}
We implant the nuclear star cluster at the center of the Milky Way-mass galaxy and study how the star cluster evolves under the tidal field of the galaxy.
The simulation is run with the finest spatial resolution of 80 pc, which is an identical choice to the AGORA project. 
In the background gravity that is computed with the PM solver, the gravity within a cell is interpolated depending on the choice of interpolation methods.
Thus, the spatial resolution of 80 pc, compared to the size of the star cluster, is apparently not sufficient to resolve the exact tides and the sophisticated physical reaction of star cluster to it, but sufficient for testing the performance of the direct N-body solver in {\newenzo} compared to {\enzo}.

\begin{figure}
    \centering
    \includegraphics[width=0.47\textwidth]{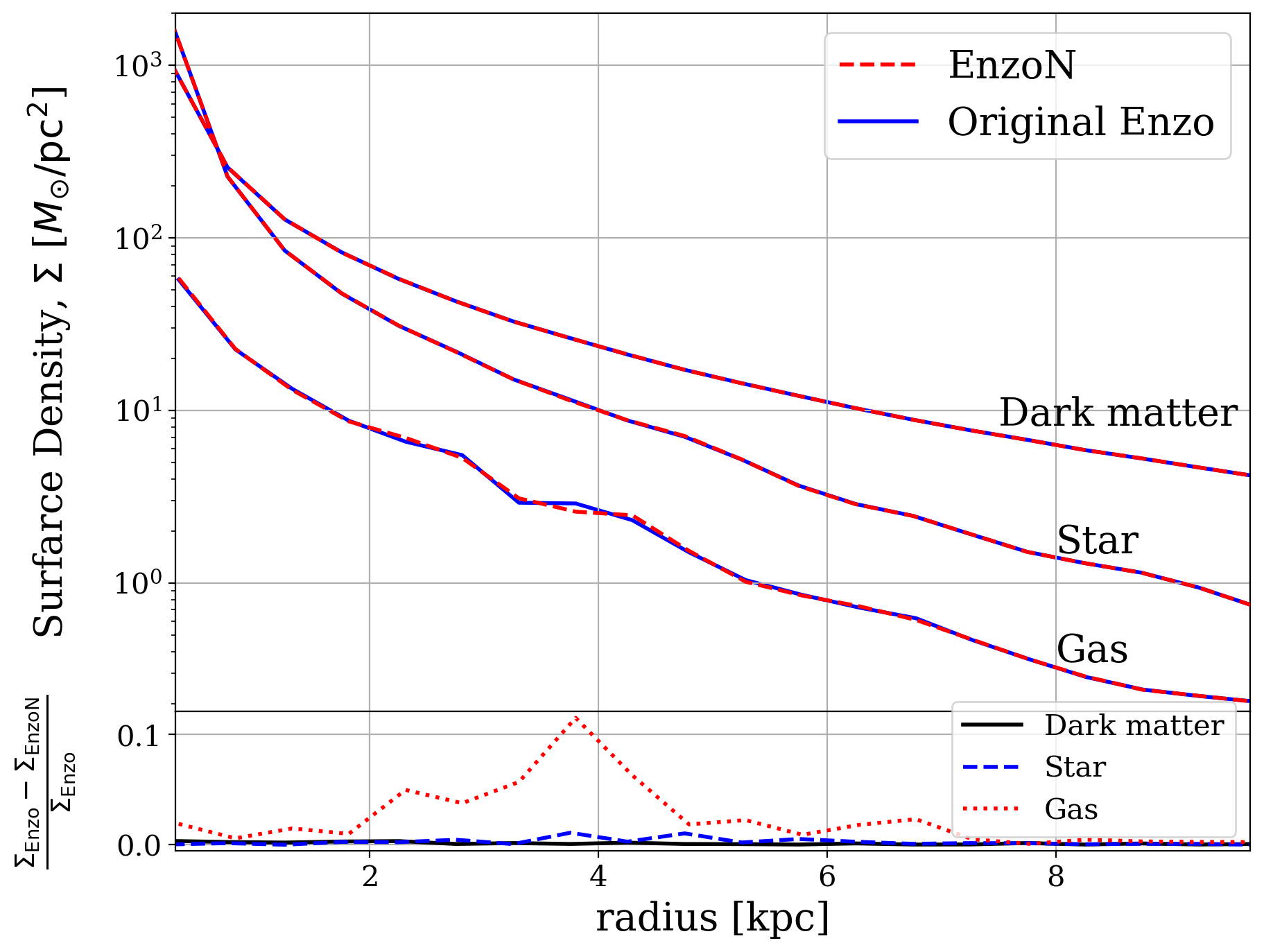}
    \caption{Comparison of the surface densities of dark matter, stars, and gas in the galaxy at $t=240\myr$ between {\newenzo} ({\it red dashed}) and {\enzo} ({\it blue solid}).
    The {\it bottom} panel displays the relative deviations between {\enzo} and {\newenzo} for dark matter ({\it black solid}), stars ({\it blue dashed}), and gas ({\it red dotted}) .
    Note that the term ``star'' in this plot refers to the stars within the galaxy, excluding those in the star clusters.
    }
    
    \label{fig:agora_nsc_profile}
\end{figure}

Fig. \ref{fig:agora_nsc} illustrates the evolution of the galaxies and their nuclear star clusters in {\enzo} ({\it top}) and {\newenzo} ({\it bottom}). 
The {\it left} panels displays the gas density projections at $t=0$, 120, and 240 Myr on a 24 kpc window. 
Since the gravitational influence from the star cluster of $10^5 \msun$ is relatively negligible compared to that of the host galaxy of $\sim 10^{12} \msun$, the overall evolution of the galaxies shows identical between {\newenzo} and {\enzo} and has not deviated from the original AGORA run having no star cluster as well (refer to Fig. 2 of \citet{kim2016ApJ...833..202K}).
This is also evident in Fig. \ref{fig:agora_nsc_profile}, which displays surface densities of dark matter, stars, and gas at $t=240\myr$. 
The surface densities of each component in {\newenzo} and {\enzo} exhibit a close agreement with negligible errors.

In contrast, the clusters scale view ({\it right} panels) of Fig. \ref{fig:agora_nsc} display significant discrepancies between {\newenzo} and {\enzo}.
The star cluster of {\enzo} is scattered, and the surface density of the star cluster at its core diminishes to $\sim 10^2 \msun/\pc^2$. 
In the meantime, the core surface density of {\newenzo} sustains its value of $\sim 10^3 \msun/\pc^2$ consistently throughout the evolution following the Plummer profile.
As discuss in Sec. \ref{sec:iso_enzon_enzo}, the disruption of the star cluster in {\enzo} is attributed to the lack of resolution.
However, the star cluster in {\enzo} is not entirely destroyed due to the gravity from surrounding matter such as dark matter and the stellar bulge, providing an extra in-falling gravity on top of self-gravity of the star cluster.  
Despite the gravitational support, it is not sufficient to maintain its structural integrity, and the star cluster is dominated by and suffers from the numerical artifact, which leads to the distortions of the morphology.

\begin{figure}
    \centering
    \includegraphics[width=0.47\textwidth]{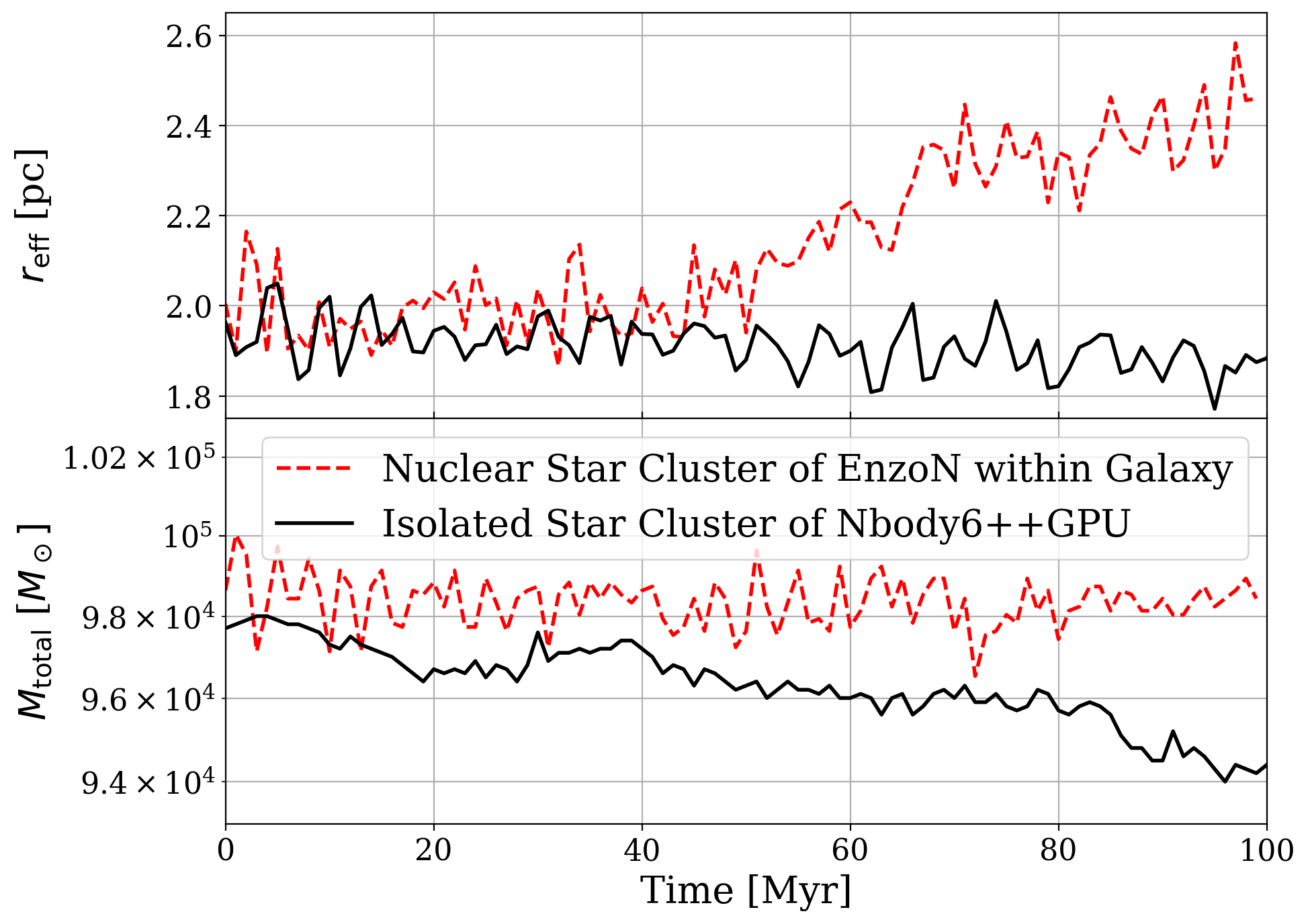}
    \caption{
     {\it Top}: Effective radius of the nuclear star cluster within the galaxy in {\newenzo} ({\it red dashed}) and the isolated star cluster in {\nbody} ({\it black solid}) as a function of time.
    {\it Bottom}: Total mass of the clusters as a function of time.
    The total mass is defined as the enclosed mass within a radius at which the surface density equals the threshold value of $1 \msun/\pc$.
    }
    
    \label{fig:agora_nsc_mass_radius}
\end{figure}

The gravitational support, namely tidal fields, should have an impact on the {\newenzo}'s star cluster as well.
Shown in Fig. \ref{fig:agora_nsc_mass_radius} is the comparison between the isolated star cluster of {\nbody} ({\it black solid}) and the nuclear star cluster of {\newenzo} ({\it red dashed}) in terms of effective radius ({\it top}) and the total mass of cluster ({\it bottom}). 
In contrast to the orbiting star cluster in Sec. \ref{sec:orbit_circular}, the nuclear star cluster does not undergo mass loss but instead sustains its mass more effectively than the isolated star cluster.
In the meantime, the size of the star cluster increases, inevitably leading to a more core profile in the center. This can be attributed to the injection of thermal energy from tidal fields.

We demonstrate that {\newenzo} can robustly handle a star cluster in the context of galaxy evolution.
The nuclear star cluster in {\newenzo} co-evolves properly with its host galaxy, maintaining the integrity of internal dynamics of the star cluster.
Furthermore, a noteworthy feature unique to {\newenzo} is that, through few-body interactions facilitated by the regularization (see Sec. \ref{sec:method_regularization}), a fraction of the particles initially associated with the star cluster can be observed in other regions of the galaxy ({\it lower} panels of the {\it right} plot in Fig. \ref{fig:agora_nsc}), which is a major missing part in {\enzo}. 
This capability opens up new avenues for investigating physical phenomena occurring on very small scales within the broader context of galactic dynamics.


\begin{figure*}
    \centering
    \vspace{-20mm}
    \includegraphics[width=1.\textwidth]{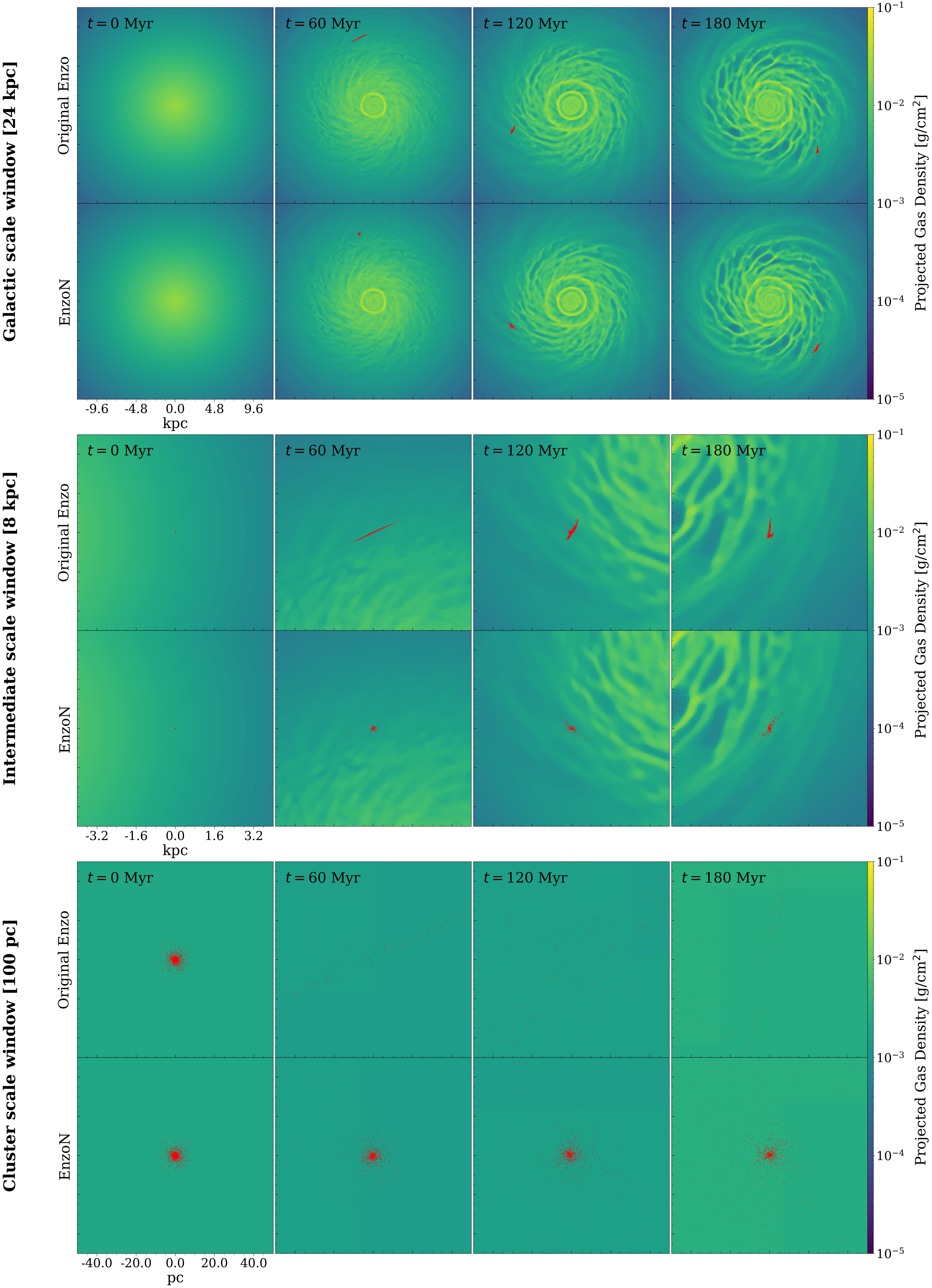}
    \caption{
    Gas and particle projections of the isolated galaxy with the star clusters on the outskirts in {\enzo} ({\it upper}) and {\newenzo} ({\it lower}).
    The star clusters ({\it red dots}) are implanted at $x=8\kpc$ in the AGORA galaxy \citep{kim2016ApJ...833..202K} (refer to Fig. \ref{fig:agora_nsc}).
    {\it Top}: Galactic scale view exhibits the similarity of gas patterns in {\enzo} and {\newenzo}. 
    {\it Middle}: Intermediate zoomed-in view manifests how differently star cluster in {\enzo} and {\newenzo} are affected by the tidal forces.
    {\it Bottom}: Cluster scale view shows the star cluster entirely disrupted in {\enzo} in contrast to the intact nuclear star cluster in {\newenzo}.
    }
    \label{fig:agora_8kpc}
\end{figure*}

\begin{figure}
    \centering
    \includegraphics[width=0.47\textwidth]{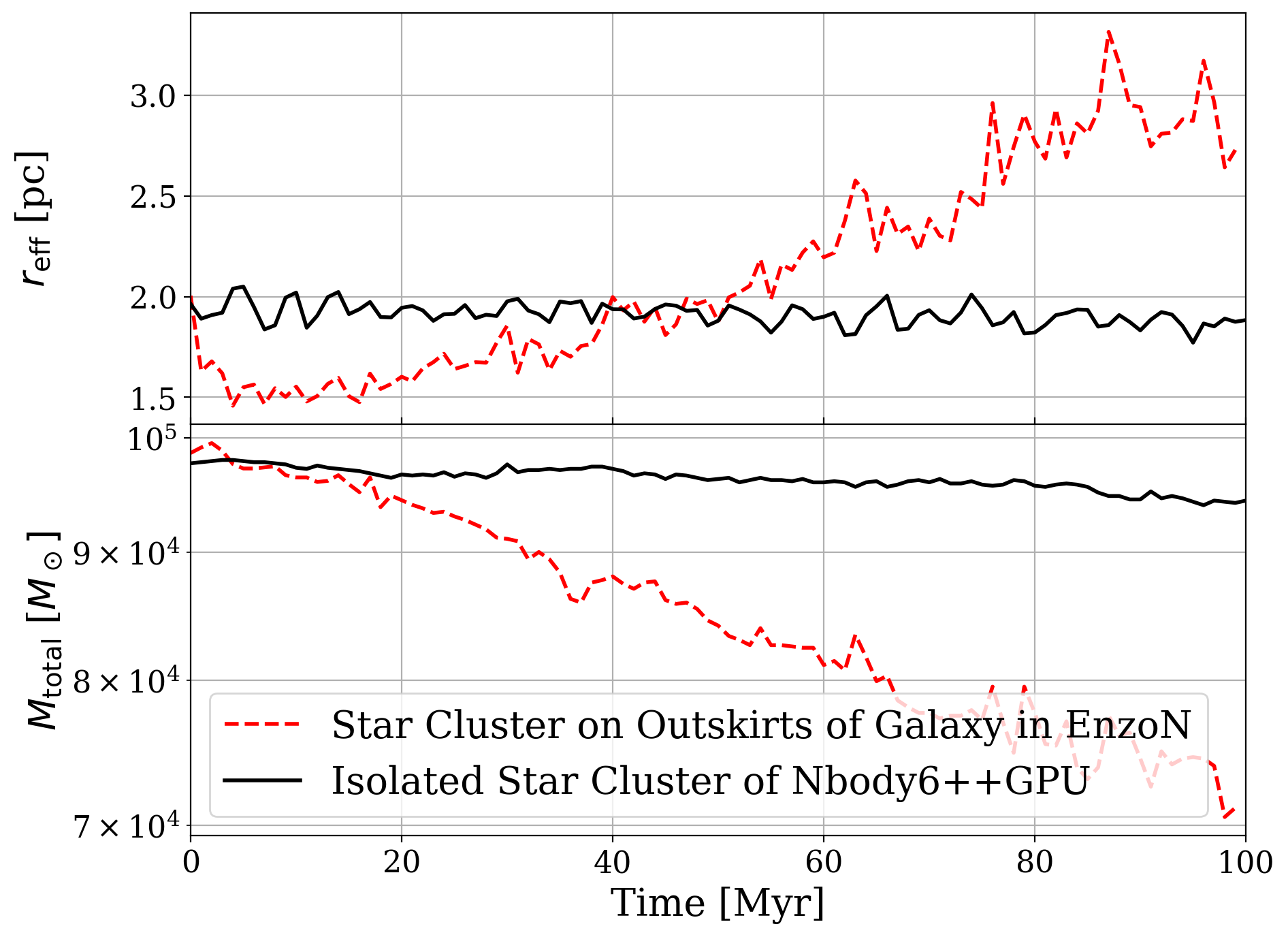}
    \caption{
    {\it Top}: Effective radius of the orbiting star cluster on the outskirts of the galaxy in {\newenzo} ({\it red dashed}) and the isolated star cluster in {\nbody} ({\it black solid}) as a function of time.
    {\it Bottom}: Total mass of the clusters as a function of time.
    The total mass is defined as the enclosed mass within a radius at which the surface density equals the threshold value of $1 \msun/\pc$.
    }
    \label{fig:agora_outskirts_mass_radius}
\end{figure}

\subsection{Star Cluster Orbiting on the Outskirts of the Galaxy}
\label{sec:agora_outskirts}
In this section, we investigate the evolution of a star cluster orbiting on the outskirts of the galaxy.
We situate the star cluster at $x=8\kpc$ at $t=0\myr$ with the same rotational velocity as that of the galaxy, with initial conditions and refinement scheme identical to those in Sec. \ref{sec:agora_nsc}.
Exhibited in Fig. \ref{fig:agora_8kpc} are the gas projections with star clusters ({\it red dots}) on three different scales: galactic scale ({\it top}), intermediate scale ({\it middle}), and cluster scale ({\it bottom}).
The results from the original {\enzo} ({\it upper}) and {\newenzo} ({\it lower}) are juxtaposed in each scale plot for comparison.
On the galactic scale, the structures of the galaxies in both {\newenzo} and {\enzo} show negligible discrepancies, while the morphology of the star clusters already exhibits notable differences.
This distinction becomes more evident in the intermediate scale window.
In {\enzo}, the particles initially constituting the star cluster are stripped along the tidal field at $t=60\myr$ and transform into an arbitrary structure at $t=120$ and $180 \myr$, indicating numerical artifacts.
This primarily originates from the lack of self-gravity due to limited resolution and the gravitational tidal forces exerted by the galaxy.

Nonetheless, the tidal field has an impact not only on the star cluster of {\enzo} but also on that of {\newenzo}.
Yet, the manifestations of stripping and morphological distortion in {\newenzo} clearly differ from those in {\enzo}.
At $t=60\myr$, the star cluster of {\newenzo} exhibits only minimal dispersion.
Subsequently, at $t=120$ and $180\myr$, traces of tidal stripping become evident along the tidal fields, yet the structure of the star cluster is maintained in contrast to {\enzo}.
This is more evident in the cluster-scale view, demonstrating that the star cluster in {\newenzo} sustains its shape throughout the time period with some dispersion due to the tides, whereas that of {\enzo} is demolished to the extent that we cannot even discern the trace of the star cluster.

Figure \ref{fig:agora_outskirts_mass_radius} depicts the impact of tidal stripping on the effective radius and total mass of the star cluster in {\newenzo} over time. 
Contrasted with the isolated star cluster of {\nbody} (black solid line), the star cluster orbiting on the outskirts of the galaxy (red dashed line) experiences a reduction in total mass and an increase in size owing to the tidal field of the parent galaxy.
This leads to a reduced core density and a dispersed profile, contrasting with the behavior of the star cluster orbiting around the massive point particle where the size of the star cluster shrinks.
This is overall in line with the results of \citet{fujii2008ApJ...686.1082F} who perform self-consistent N-body simulations containing a star cluster and its parent galaxy. 
However, conducting a detailed analysis is challenging due to the increased complexity of the galactic tidal fields compared to the more ideal and analytic tidal fields discussed in Sec. \ref{sec:orbit_circular}.
Nonetheless, it can be speculated that the energy injected by the tidal force is comparable to the binding energy of the star cluster, preventing core collapse and leading to gradual dispersion of the star cluster.

We have conducted a simulation of a star cluster orbiting on the outskirts of a galaxy as a potential scientific application.
{\newenzo} demonstrates the capability to successfully simulate both the evolution of the galaxy and the star cluster, capturing its internal dynamics.
The evolution of the star cluster within the galaxy distinctly differs from both isolated and ideal orbiting cases, emphasizing the necessity of {\newenzo} for studying star clusters in complex backgrounds. 
In comparison to {\nbody}, {\newenzo} can nurture star clusters in various more realistic environments, such as inside a galaxy, in the vicinity of a galaxy, or within the larger-scale structure.
{\newenzo} opens up new avenues for intriguing scientific inquiries, such as the formation and evolution of star clusters within a galaxy, in-situ and ex-situ star formation history, and the migration of globular clusters into a galaxy.

\section{Summary}
\label{sec:summary}
In this study, we introduce a novel hybrid framework, denoted as {\newenzo}, designed to enable robust self-consistent simulations of star clusters within their parent galaxies.
This involves the seamless integration of the direct N-body code, {\nbody}, into the (magneto-)hydrodynamic code, {\enzo}.
{\nbody} is purpose-built for simulating star clusters, utilizing a direct N-body approach, fourth-order Hermite integrator, and KS regularization (see Sec. \ref{sec:method_nbody}). 
Meanwhile, {\enzo} is designed to address a wide array of astrophysical problems, employing various physics engines such as (magneto-)hydrodynamics, PM gravity solver, gas chemistry, radiative cooling, cosmological expansion, and sub-resolution models for star formation and feedback.
However, resolving the highly nonlinear and chaotic internal dynamics, crucial for understanding star clusters, has proven challenging due to the absence of a higher-order time stepper, and limitations of the PM gravity solver in {\enzo}.
Conversely, the limited range of physics implemented in {\nbody} has restricted the study of star clusters in diverse environments interacting with other astrophysical components.

In {\newenzo}, we have tackled this challenge by linking two distinct frameworks through semi-stationary background acceleration of the `hydro' part.
This address the compatibility issue in the time stepping between the `hydro' part and the `nbody' part.
The background acceleration is computed from dark matter, gas, and stars---entities not subject to direct N-body interactions---on the hydro part. 
Subsequently, the nbody part utilizes this background acceleration as tidal fields. 
It calculates collisional gravity for the target particles in star clusters in addition to the background acceleration.
The background acceleration remains stationary during $\dt_\mathrm{hydro}$.
Every $\dt_\mathrm{hydro}$, the positions and velocity calculated in the `nbody' part are relayed to the `hydro' part, contributing to the full acceleration calculation.
Simultaneously, the background acceleration is updated on the `hydro' part and passed to the `nbody' part for the direct N-body calculation of the next time step (refer to Sec. \ref{sec:method_hermite} for details).

We have conducted a series of tests to examine the robustness and liability of {\newenzo}.
The objectives of the tests are the following: 
1. ensuring the integrity of the implemented direct N-body calculation (refer to Sec. \ref{sec:isolated_star_cluster});
2. verifying the robustness of the background acceleration (refer to Sec. \ref{sec:orbit});
3. exploring the potential for synergetic scientific applications (refer to Sec. \ref{sec:agora_star_cluster}).
The test results are summarized as follows:
\begin{itemize}
    \item The implementation of the direct N-body solve in {\newenzo} agrees well with {\nbody} in an isolated star cluster test (refer to Sec. \ref{sec:iso_enzon_nbody6}).
    The evolution of star clusters in {\newenzo} coincide with {\nbody} within the confidence level of $1\sigma$ (see Fig. \ref{fig:lagr_nb_enzon}). 
    In the primordial binary test, the evolution of the binary fractions in {\newenzo} and {\nbody} are in a good agreement within only a few percent errors, suggesting that the integrity of the regularization, handling few-body interactions (refer to Sec. \ref{sec:method_regularization}), is robust (see Fig. \ref{fig:bin_hist}).
    
    \item In the test where a massive particle is placed at the center and the background acceleration is measured, the background acceleration closely follows the analytic solution with an error less than 0.1\% (see Fig. \ref{fig:background_analytic} in Sec. \ref{sec:orbit_analytic}).
    Further, the star cluster, situated near the massive particle, successfully follows an orbit around the primary with a relative error of 1.32\% in a radial orbit and less than 1\% error in the period, while maintaining the structural integrity of the star cluster for 100 Myr (refer to Sec. \ref{sec:orbit_circular}). 
    
   \item In Sec. \ref{sec:agora_star_cluster}, the star cluster is implanted at the center and on the outskirts of the galaxy, respectively. 
   The nuclear star cluster can evolve under the central tidal fields, which bind the star cluster additionally, resulting in a decline in the total mass and size of the cluster (refer to Sec. \ref{sec:agora_nsc}).
   In the case of the star cluster orbiting on the outskirts of its parent galaxy, it experiences tidal stripping by the galactic tides but manages to persist during 180 Myr (refer to Sec. \ref{sec:agora_outskirts}).
   A caveat is that tidal fields may contain numerical artifacts to some extent due to the relatively low resolution of the simulations.
\end{itemize}

In addition to the robustness and liability tests, we also performed a comparative analysis between the results obtained from {\newenzo} and those from {\enzo} for each of the aforementioned tests above.
The comparative analysis can be succinctly summarized.
In {\enzo}, maintaining the structure of star clusters requires a resolution of at least several parsecs.
Even with an extremely high resolution (e.g., 0.08 pc), star clusters in {\enzo} fail to evolve over time, barely sustaining their structures (see Fig. \ref{fig:lagrangian_radius_enzo_enzon}).
In contrast, {\newenzo} can consistently simulate the star cluster regardless of resolution, offering the identical physical precision to {\nbody}.
This capability significantly reduces the computational time when {\newenzo} and {\enzo} aim to star clusters.
For instance, {\newenzo} with a resolution of $20 \pc$ is eleven times faster, even with higher accuracy, than {\enzo} with a resolution of $0.08 \pc$ in the case of the star cluster orbiting the massive point particle (refer to Sec. \ref{sec:orbit_circular}).
Even with the identical refinement scheme, where only {\newenzo} has the computational disadvantage of computing the additional `nbody' part, the penalty in the computational time approximates to only $\sim 10\%$ of {\enzo} in general.

In conclusion, {\newenzo}, designed to simulate star clusters in various astrophysical environments, successfully integrates the strengths of both {\enzo} and {\nbody}.
Specifically, {\bf 1) {\newenzo} can simulate star clusters with the precision of {\nbody} unaffected by the resolution of hydrodynamics};
{\bf 2) {\newenzo} excels at simulating any scientific environment that {\enzo} can handle, seamlessly incorporating star clusters without significant additional computational overhead.}

\section{Future work}
{\newenzo} demonstrates significant potential for exploring the evolution of star clusters across diverse contexts.
Nonetheless, its current version necessitates seamless integration of various physics for further enhancement.
In the following studies, our focus will be on implementing star formation and feedback, cosmological evolution, black hole formation and feedback, and developing an on-the-fly star cluster finder. 
This comprehensive approach will enable us to delve into the formation and evolution of the first star cluster within the context of cosmology in the early universe.
Moreover, our study will extend to investigating the formation of intermediate mass black holes in star clusters via runaway collisions and its evolution in a cosmological simulation.
This has been examined through observational studies \citep{zwart}, semi-analytical approaches \citep{devecchi2009ApJ...694..302D}, and even idealized simulations \citep{das2021MNRAS.503.1051D}.
However, its exploration within the framework of self-consistent cosmological simulations has remained unaddressed and challenging.

\acknowledgments
The Flatiron Institute is supported by the Simons Foundation.
Ji-hoon Kim’s work was supported by the National Research Foundation of Korea (NRF) grant funded by the Korea government (MSIT) (No. 2022M3K3A1093827 and No. 2023R1A2C1003244). His work was also supported by the National Institute of Supercomputing and Network/Korea Institute of Science and Technology Information with supercomputing resources including technical support, grants KSC-2020-CRE-0219, KSC- 2021-CRE-0442 and KSC-2022-CRE-0355.
GLB acknowledges support from the NSF (AST-2108470, XSEDE grant MCA06N030), a NASA TCAN award 80NSSC21K1053, and the Simons Foundation (grant 822237) and the Simons Collaboration on Learning the Universe.
The publicly available Enzo and yt code used in the analysis of this work is the product of collaborative efforts by many independent scientists from numerous institutions around the world. Their commitment to open science has helped make this work possible.

\bibliography{main}{}
\bibliographystyle{aasjournal}

\end{document}